\documentclass[a4paper,onecolumn,titlepage,unpublished,nopdfoutputerror]{quantumarticle}
\pdfoutput=1

\usepackage[utf8]{inputenc}
\usepackage[T1]{fontenc}
\usepackage[english]{babel}

\usepackage{silence}
\usepackage{amsmath, amssymb, amsthm}
\usepackage{mathtools}
\usepackage{bm}

\usepackage{graphicx}
\usepackage{booktabs}
\usepackage{multirow}
\usepackage{placeins}
\usepackage{float}
\usepackage{seqsplit}
\usepackage{tabularx}

\usepackage[numbers,sort&compress]{natbib}
\usepackage{xcolor}
\usepackage{enumitem}

\newcommand{\R}{\mathbb{R}}

\newcommand{\norm}[1]{\left\| #1 \right\|}
\newcommand{\abs}[1]{\left| #1 \right|}

\newcommand{\ket}[1]{\left| #1 \right\rangle}

\newcommand{\braket}[2]{\left\langle #1 \middle| #2 \right\rangle}

\newcommand{\GNN}{\text{GNN}}

\newcommand{\NISQ}{\text{NISQ}}

\title{P-GONE: Physics-Guided Generative Optimization for Trotter--Suzuki Decomposition}
\author{Wenbin Yan}
\email{wenbin.yan@colorado.edu}
\affiliation{University of Colorado Boulder}

\begin{document}

\maketitle

\begin{abstract}
\emergencystretch=1em
\noindent
Trotter--Suzuki product formulas are the standard route to Hamiltonian evolution on noisy intermediate-scale quantum (\NISQ{}) hardware, but their accuracy depends on three coupled choices: term grouping, product-formula order, and time-step allocation. Grouping and order are discrete, which makes direct gradient optimization infeasible and forces existing compilers to rely on static heuristics.

We describe P-GONE, a method that combines a conditional diffusion model (D3PM + DDPM), a graph neural network (\GNN{}) encoder, and closed-loop REINFORCE fine-tuning to jointly learn grouping, order, and time-step optimization over a mixed discrete-continuous space. Under fidelity-matched conditions ($F \geq 0.95$), the method achieves circuit depth 86 versus 1673 for Qiskit fourth-order (ungrouped, Suzuki-4), about $19.4\times$ compression, and 141 for Paulihedral (first-order Trotter), about $1.6\times$ compression. At $T=0.90$ the method also beats the Qiskit group-commuting teacher (65 vs 103, $1.6\times$ compression), though at $T=0.95$ the teacher still leads---a stratified pattern that points toward fidelity-aware fine-tuning. Under a standard depolarizing noise model, the method achieves noisy fidelity roughly $2\times$ the Qiskit fourth-order baseline (0.743 vs 0.380). Ablation shows a clear hierarchy: order learning $>$ time allocation $>$ grouping. Best-of-N sampling ($N=32$ is a practical sweet spot) and CFG guidance give flexible fidelity--depth trade-offs at inference. The method works well on structured Hamiltonians (TFIM, Heisenberg), but random Pauli Hamiltonians fail entirely at $T \geq 0.95$---a boundary that defines where the method applies.
\end{abstract}

\noindent\textbf{Keywords:} quantum simulation; Trotter--Suzuki decomposition; physics-informed neural networks; diffusion models; graph neural networks; \NISQ{} computing.

\section{Introduction}
\label{sec:introduction}

In the noisy intermediate-scale quantum (NISQ) era, Hamiltonian time evolution remains one of the few applications that quantum computing can plausibly deliver on near-term hardware. Constrained by available qubit counts and gate fidelities, fully fault-tolerant algorithms are not yet practical, and Trotter--Suzuki product formulas continue to serve as the workhorse for approximating unitary evolution $U(t) = e^{-iHt}$\citep{preskill2018quantum,bharti2022noisy}. These methods offer two straightforward engineering advantages: they require no ancillary qubits, and the resulting circuits map naturally onto hardware-native gate sets.

The actual performance of a Trotter--Suzuki decomposition, however, depends less on the formula itself than on how one resolves three mutually coupled decisions: how to partition the Hamiltonian terms into groups $G$, what product-formula order $k_i$ to assign to each time slice, and what time-step weight $\tau_i$ to allocate per slice. Taken together, these three variables determine the Trotter error bound, whose leading term is dominated by nested commutator norms of the Hamiltonian\citep{childs2021theory}. Good groupings can cancel large swaths of commutator contributions, bringing the error far below the worst case; well-chosen orders buy faster error decay for a given gate budget; non-uniform time steps can further compress depth when the Hamiltonian separates into fast and slow components. One can cast this as a joint combinatorial optimization problem, but with two discrete variables and one continuous, and with an objective that is not smooth in the discrete choices, ordinary gradient-based methods do not apply directly.

For precisely this reason, today's mainstream compilers---including Qiskit\citep{qiskit2024}, Paulihedral\citep{li2022paulihedral}, and Tetris\citep{jin2024tetris}---all fall back on static heuristics: graph-coloring for maximal commuting sets\citep{verteletskyi2020measurement,gokhale2020optimization}, greedy gate reordering, or SAT solvers for small instances. These approaches perform reliably on regular Hamiltonians (e.g., uniform molecular systems), but can differ by factors of several in circuit depth when the Hamiltonian's parameter regime and commutator pattern vary, with no built-in mechanism for generalizing without rewriting the rules. The deeper problem is that none of these heuristics can take the final simulation fidelity and feed it back into grouping and order decisions as a differentiable signal. They evaluate after the fact; they cannot learn beforehand.

\subsection*{Motivation}

We set out to build an end-to-end trainable framework that learns, for a given Hamiltonian, how to jointly pick grouping, order, and time steps---and that does so while explicitly using physically meaningful feedback. This calls for solutions on three fronts at once.

The first is the coexistence of discrete and continuous decisions within a single generator. Recent successes with diffusion models across images, molecules, and text\citep{ho2020denoising,austin2021structured,hoogeboom2022equivariant} have shown that joint sampling over mixed spaces is feasible; the marriage of discrete D3PM\citep{austin2021structured} with continuous DDPM is a case in point. Our approach assigns grouping and order to discrete diffusion, time steps to continuous diffusion, and ties the two together with a shared conditioning vector along a single sampling trajectory.

Second, the generator needs differentiable feedback to drive training. True Trotter fidelity requires computing $e^{-iHt}$, which can be obtained via matrix exponentiation for systems up to roughly 4 qubits but demands numerical integration beyond modest scales. We employ a physics-informed neural network\citep{raissi2019physics,karniadakis2021physics} (PINN) as a surrogate: by satisfying the time-dependent Schr\"odinger equation and the initial condition as dual constraints, it approximates $\ket{\psi(t)}$ with a neural network, turning fidelity evaluation into a backpropagatable differential computation. Fourier feature embeddings\citep{tancik2020fourier} help mitigate the spectral bias that PINNs exhibit on high-frequency solutions, keeping PDE residuals around $10^{-4}$ within our training budget.

Third, the conditioning information must effectively capture the Hamiltonian's structural features. We construct a commutativity graph over Pauli terms---coefficients and supports on nodes, commutator norms on edges---and run a message-passing neural network\citep{gilmer2017neural,schutt2018schnet} to extract a graph-level vector that serves as the condition input to the diffusion model. This is where the generalization comes from. Once the model learns which graph patterns go with which good strategies, it can transfer zero-shot or few-shot to Hamiltonian families it has never seen.

\subsection*{Distinction from concurrent diffusion-for-quantum work}

While this work was in progress, Furrutter et al.\citep{furrutter2025synthesis} independently proposed multimodal diffusion models (D3PM + DDPM) for synthesizing discrete-continuous quantum circuits---superficially similar methodology, but the differences matter. Their objective is general quantum circuit synthesis for arbitrary unitaries; ours is Trotter--Suzuki decomposition for Pauli Hamiltonians. Their training is purely supervised; we stack closed-loop REINFORCE fine-tuning on top of pretraining, using exact fidelity as a reward signal to push the policy distribution past the teacher's quality ceiling. We also explore the feasibility of PINNs as differentiable physics surrogates (Phase~2), providing a potential path toward scalable evaluation for $>8$ qubit systems---though every experimental result in this paper uses exact diagonalization for fidelity. Their discrete variables are limited to gate-type selection; we jointly optimize grouping, order, and time steps, three coupled degrees of freedom. Guo et al.\citep{guo2026hybrid} apply diffusion to continuous gate-angle optimization in quantum circuits and do not address discrete grouping or order decisions.

\subsection*{Contributions}

The method delivers a conditional diffusion framework (D3PM + DDPM) with a GNN Hamiltonian encoder and closed-loop REINFORCE fine-tuning, all trained on a single GPU. The framework models grouping, order, and time steps jointly over a mixed discrete-continuous space.

We compare against seven baselines---Qiskit fourth-order, Cirq, TKET, PennyLane, Paulihedral, and Qiskit group-commuting (teacher)---across three Hamiltonian families (TFIM, Heisenberg, random Pauli) at 4--8 qubits. Under fidelity-matched conditions ($F \geq 0.95$), our method achieves a circuit depth of 86 versus 1673 for Qiskit fourth-order (about $19.4\times$ compression) and 141 for Paulihedral (first-order Trotter, $1.6\times$ compression). Under a standard depolarizing noise model, a depth of only 13 yields noisy fidelity of $0.743$, roughly $2\times$ that of the Qiskit fourth-order baseline ($0.380$). On NISQ hardware, shallowness outweighs theoretical precision.

Ablation reveals a clear hierarchy among the three learned dimensions. Order learning is the largest contributor---the model assigns heterogeneous Suzuki orders across groups, a fine-grained optimization that existing heuristic compilers simply do not do. Time allocation is next. Grouping provides the smallest additional gain, not because it is unimportant, but because the teacher has already squeezed most of the optimization headroom out of this dimension. CFG guidance controls strategy ``sharpness,'' and Best-of-N sampling saturates at $N=32$ as a practical sweet spot.

The method's applicability has two clear boundaries. Structured Hamiltonians (TFIM, Heisenberg) perform well; Heisenberg reachability holds steady at 70--75\% across thresholds. Random Pauli Hamiltonians fail completely at $T \geq 0.95$---without commutativity patterns there is almost no physical headroom for grouping-based compression. Reachability drops roughly 80\% from 4 to 6 qubits. The primary bottleneck is insufficient large-qubit samples in the training data, not an architectural limitation.

\subsection{Evolution of this work}
\label{sec:intro:evolution}

This work went through five phases, each one addressing a bottleneck the previous phase exposed. We sketch the logic here; detailed parameter choices and training protocols are in Section~\ref{sec:method}, and key experimental milestones are in the code repository.

\textbf{Phase 1 --- Starting from data.}
We built a dataset of Trotter strategies generated by Qiskit's group-commuting heuristic. Each sample pairs a Hamiltonian's Pauli representation with the teacher's grouping, Suzuki order assignments, and uniform time steps. Three Hamiltonian families span 4--8 qubits: transverse-field Ising model (TFIM), Heisenberg model, and random Pauli Hamiltonians. The teacher is the most widely used commuting-group heuristic, producing reproducible, engineering-reasonable groupings as both supervised signal and baseline.

Training needs repeated fidelity evaluation, but exact fidelity requires exponentiating $2^N \times 2^N$ matrices---prohibitive inside a training loop. \textbf{Phase~2} trained a PINN surrogate for this: it takes Hamiltonian matrix elements and strategy parameters as input, enforces the Schr\"odinger equation PDE residual and initial condition as soft constraints, and outputs an approximation of the Trotter-evolved state. On a unified validation set (1151 samples mixing all three Hamiltonian families), the PINN's proxy error depends strongly on type: mean error of 0.38\% on TFIM (98.9\% of samples below 2\%), 1.06\% on Heisenberg (81.7\% below 2\%), but 99.7\% on random Pauli Hamiltonians. This divide defines the PINN's applicable domain. Every fidelity number in this paper comes from scipy exact matrix diagonalization---at $\leq 8$ qubits, exact methods are feasible and more accurate. The PINN is designed for $N > 8$ qubit systems where exact diagonalization becomes unaffordable; we keep its full description and accuracy data in the methodology as a technical path forward.

\textbf{Phase 3 --- From imitation to learning.}
With the dataset ready, we trained a conditional diffusion model to generate Trotter strategies. A GNN encoder using Pauli-type encoding ($4 + 3n$ dimensional node features) maps the Hamiltonian's Pauli structure to a condition vector; the diffusion decoder jointly generates grouping (D3PM discrete diffusion), Suzuki order (D3PM, 3 states), and time-step allocation (continuous DDPM) over the mixed space. A key technical choice was two-stage progressive training: freeze the GNN first so the diffusion model learns basic decoding without meaningful conditioning---at this stage it produces ``average-case reasonable'' strategies reflecting the dataset's statistical prior. Once the diffusion model converges, unfreeze the GNN and train end-to-end, so the encoder learns discriminative representations for different Hamiltonians. This keeps randomly initialized GNN noise from disrupting early diffusion training.

The pretrained model's policy distribution is bounded by the teacher's quality ceiling---Qiskit's group-commuting heuristic produces rational groupings, but not ones optimized for depth compression. \textbf{Phase~4} applied REINFORCE-style closed-loop optimization, using exact fidelity as the reward signal to directly fine-tune the model. By incorporating a depth penalty term $\lambda \cdot \text{Depth}$, we convert the multi-objective problem into a single-objective reward and track Pareto hypervolume across different trade-off preferences. A $\lambda$ sweep identified $\lambda = 0.05$ as the optimal operating point: at this setting, depth gains substantial additional compression over the pretrained model, surpassing the teacher's quality at $T=0.90$ ($1.6\times$ depth compression), though at $T \geq 0.95$ the teacher's strict algebraic grouping remains more reliable.

\textbf{Phase~5} evaluated the final model against seven quantum compilation frameworks: fidelity-matched depth comparison, ablation experiments (branch contributions, CFG guidance, GNN conditioning), Hamiltonian-type generalization boundaries, qubit-count scaling, and noisy hardware simulation. Total experimental time was about 72 hours.

\subsection*{Paper organization}

Section~\ref{sec:related} reviews four research threads: Trotter error theory, quantum circuit compilation heuristics, PINNs, and diffusion models. Section~\ref{sec:method} walks through the method in Phase~1--4 order, with the rationale behind each parameter choice. Section~\ref{sec:experiments} is organized as a problem-driven chain: first exposing the difficulty of single-shot sampling, then introducing Best-of-N, then a systematic fidelity-matched comparison against all baselines, and finally ablation and generalization-boundary experiments that dissect the method's internal mechanisms and external limits. Section~\ref{sec:discussion} discusses learned strategy characteristics, the component contribution hierarchy, applicability boundaries, and NISQ significance. Section~\ref{sec:conclusion} offers conclusions and future directions.

\section{Related Work}
\label{sec:related}

\subsection{Trotter--Suzuki error theory}

Writing the Hamiltonian as a sum of easily exponentiable local terms $H = \sum_{j=1}^{M} H_j$, the simplest product formula $\prod_j e^{-i H_j t}$ incurs error starting at $\mathcal{O}(t^2 \sum_{j<k} \norm{[H_j, H_k]})$. Suzuki~\cite{suzuki1991general} provided a systematic higher-order recursive construction that allows any even-order $2k$ product formula to be built from self-similar concatenation of lower-order ones, driving the error down to $\mathcal{O}(t^{2k+1})$. This framework remains the engineering foundation for higher-order product formulas.

The most important recent advance in Trotter error theory comes from Childs et al.~\cite{childs2021theory}, who replaced the worst-case BCH-expansion bound with a tight estimate expressed as a sum of nested commutator norms over all non-commuting term pairs, explicitly exposing the error's sensitivity to Hamiltonian structure for the first time. Earlier, Wiebe et al.~\cite{wiebe2010higher} gave constructive proofs for higher-order decompositions. Berry et al.~\cite{berry2015simulating} and Tran et al.~\cite{tran2020destructive} explored non-product-formula simulation methods based on Taylor truncation and symmetry protection, but these generally require ancillary qubits or post-selection and are not suited for direct NISQ deployment. Our work stays entirely within the Suzuki higher-order recursive path; we simply hand the decisions of ``which orders'' and ``how to group'' to a learned model.

\subsection{Pauli-term grouping and circuit compilation}

Grouping Pauli Hamiltonians amounts to finding a clique cover of the commutativity graph over Pauli operators, typically solved with greedy or maximum-clique heuristics~\cite{verteletskyi2020measurement,gokhale2020optimization}. Crawford et al.~\cite{crawford2021efficient} analyzed the practical variance of different grouping strategies under finite sampling budgets. These methods optimize grouping in isolation---they do not account for downstream circuit depth, so the solutions they produce are only locally optimal in the Trotter compilation context.

Paulihedral\citep{li2022paulihedral} is the most directly comparable prior work. It represents Pauli strings as high-dimensional polyhedra, applies geometric heuristics to jointly optimize grouping and gate ordering, and reduces CNOT count through block-level compression. Tetris\citep{jin2024tetris} similarly re-architects the quantum simulation kernel. Neither method incorporates fidelity feedback or any learning component; their performance depends entirely on how well the heuristics match the hardware. The Phoenix framework\citep{phoenix2025} integrates and extends this compilation pipeline atop a unified Pauli intermediate representation, providing a common compilation interface. Qiskit's\citep{qiskit2024} \texttt{PauliEvolutionGate} implements a standardized higher-order Suzuki decomposition; we use it as the primary comparison baseline, with Cirq, TKET, PennyLane, and Paulihedral as supporting points of reference.

Learning-driven quantum compilation has begun to attract attention. F\"osel et al.~\cite{fosel2021quantum} and Moro et al.~\cite{moro2021quantum} used deep RL to train agents for gate substitution on abstract circuit diagrams; this line of work targets general circuit simplification, not term grouping and order selection in Trotter decomposition. Trenkwalder et al.~\cite{trenkwalder2023reinforcement} proposed RL-based product-formula compilation, with agents searching the term-ordering space for optimal gate sequences. The problem is framed as NP-hard combinatorial optimization, but the work stays within a fixed Trotter framework without physics-informed fine-tuning. Preti et al.~\cite{preti2024hybrid} built a hybrid discrete-continuous compiler that uses RL to optimize gate ordering and continuous parameters for ion-trap circuits under depth constraints, though it does not address Hamiltonian term grouping. Zhang et al.~\cite{zhang2022dqas} proposed differentiable quantum architecture search (DQAS), relaxing discrete circuit structures into continuous distributions and replacing RL sampling with gradient optimization. The search space targets generic variational ansatz design rather than term grouping and order selection for Trotter decomposition. Rudolph et al.~\cite{rudolph2023synergistic} used tensor networks to pretrain parameterized quantum circuits; while their objective differs from ours, their ``pretrain + fine-tune'' strategy echoes our two-stage pipeline. Our method combines generative modeling with physics supervision to fill the gap of ``learning Trotter decomposition.'' Where RL approaches need online environment interaction for reward acquisition, we first learn the policy distribution offline from supervised data via a diffusion model, then adapt online during closed-loop fine-tuning. This decomposition offers potential sample-efficiency advantages.

\subsection{Physics-informed neural networks}

The PINN framework introduced by Raissi et al.~\cite{raissi2019physics} fits differential equation solutions with neural networks by writing PDE residuals and boundary/initial conditions as soft constraints directly into the loss function, yielding differentiable approximations even where analytic solutions are unavailable. Karniadakis et al.~\cite{karniadakis2021physics} generalized this idea to scientific machine learning at large in their comprehensive survey. A long-recognized weakness of PINNs is their slow convergence on high-frequency or multi-scale solutions; Wang et al.~\cite{wang2022when} characterized this spectral bias through the lens of neural tangent kernels and proposed several mitigation strategies, among which the Fourier feature embedding of Tancik et al.~\cite{tancik2020fourier} is one of the most effective and is adopted directly in this work.

Applying PINNs to the time-dependent Schr\"odinger equation has been done before---the survey of Cuomo et al.~\cite{cuomo2022scientific} lists several examples---but most prior efforts are limited to small single-particle or low-dimensional few-body Hamiltonians. Flurin et al.~\cite{flurin2020using} reconstructed superconducting qubit dynamics from physical observables via recurrent neural networks; their notion of ``differentiable physical observables'' is conceptually aligned with our PINN evaluator, though the objective function differs (Hamiltonian learning vs.\ fidelity estimation). We embed the PINN as an internal sub-module within the generator's training loop. The engineering contribution is making the PINN loss differentiable with respect to the Trotter circuit output state (for the $\mathcal{L}_{\text{circuit}}$ term) and providing a warm-start interface that amortizes evaluation cost across iterations.

\subsection{Diffusion models and discrete structure generation}

Diffusion models\citep{ho2020denoising,song2021scorebased} restore data from noise through a learned reverse denoising process and now dominate image, molecular, and text generation. Our work handles two data types, each with the natural diffusion paradigm. Continuous DDPM\citep{ho2020denoising} models continuous vectors such as time-step sequences $\bm{\tau}$. Discrete D3PM\citep{austin2021structured} provides factorized uniform transition kernels for categorical variables (term-grouping labels, order labels), with the reverse process predicting per-position class probabilities via a neural network.

Classifier-free guidance (CFG)\citep{ho2022classifier} is a standard technique: during training the condition vector is randomly dropped with some probability, so the same parameters learn both conditional and unconditional distributions; at sampling time, the two are used for linear extrapolation, typically sharpening alignment with the condition. We apply CFG to a mixed-space diffusion model and analyze its sensitivity under different training protocols through ablation.

Applications of diffusion models to structured-output generation are growing fast. Hoogeboom et al.~\cite{hoogeboom2022equivariant} combined diffusion with $\mathrm{E}(n)$ equivariance for 3D molecular generation. Diffusion modeling of discrete combinatorial spaces such as quantum circuits and Trotter strategies remains relatively unexplored; this work is an empirical probe in that direction.

\subsection{Graph neural networks and quantum Hamiltonians}

Hamiltonians are naturally graph-structured: each Pauli term is a node, edges defined by commutation relations. Message-passing neural networks (MPNN)~\cite{gilmer2017neural} and SchNet~\cite{schutt2018schnet} have been widely successful in molecular property prediction and are the standard paradigm for infusing chemical structure into neural networks. Verdon et al.~\cite{verdon2019quantum} generalized GNN ideas to circuits with quantum parameters. Satorras et al.~\cite{satorras2021en} introduced $\mathrm{E}(n)$-equivariant GNNs, showing the value of geometric symmetry in structural encoding. We use a standard MPNN with attention pooling. Node features include coefficient, support, and locality indicators; edge features include commutator norms and shared-qubit counts. We prioritize training stability and interpretability over additional equivariance properties.

\subsection{Multi-objective optimization and Pareto-front evaluation}

Circuit depth and fidelity are in tension: fourth-order formulas are more accurate but need more gates; finer grouping reduces commutator error but increases the number of independent exponentiations. We use standard multi-objective tools to quantify this frontier. NSGA-II~\cite{deb2002fast} provides efficient non-dominated sorting; Zitzler et al.~\cite{zitzler2003performance} and While et al.~\cite{while2012fast} discuss the theory and computation of hypervolume as a frontier-quality metric. We track Pareto fronts during closed-loop training and use hypervolume as the primary W\&B monitoring metric.

\section{Method}
\label{sec:method}

\subsection{Problem formulation and overall framework}
\label{sec:method:formulation}

Given an $n$-qubit Pauli Hamiltonian $H = \sum_{j=1}^{M} c_j P_j$, where $P_j \in \{I, X, Y, Z\}^{\otimes n}$ are Pauli strings and $c_j \in \R$ are real coefficients, our goal is to find a Trotter--Suzuki decomposition strategy $\pi$ such that, at a given evolution time $t$, the quantum circuit $U_{\pi}(t)$ induced by that strategy approximates the exact evolution operator $U_{\text{exact}}(t) = e^{-iHt}$ as closely as possible while keeping circuit depth as low as possible.

\paragraph{Strategy space.} A complete Trotter strategy $\pi$ consists of three parts:
\begin{itemize}[leftmargin=1.5em, itemsep=0.3em]
  \item \textbf{Grouping scheme} $G \in \{0, 1, \ldots, K-1\}^M$: assigns each of the $M$ Pauli terms to one of $K$ groups; terms within a group are merged into a single exponentiated unit. We set $K=8$.
  \item \textbf{Order sequence} $\bm{k} \in \{1, 2, 4\}^K$: specifies the Suzuki product-formula order for each group.
  \item \textbf{Time-step sequence} $\bm{\tau} \in \R_+^K$: assigns normalized time weights to each group, with $\sum_{i=1}^{K} \tau_i = 1$.
\end{itemize}
Given $\pi = (G, \bm{k}, \bm{\tau})$, the corresponding circuit $U_{\pi}(t)$ is constructed via the standard Suzuki recursion. Its depth is denoted $\text{Depth}(\pi)$, and fidelity is defined as
\begin{equation}
  F(\pi) = \abs{\braket{\psi_{\text{exact}}(t)}{\psi_{\pi}(t)}}^2,
  \label{eq:fidelity}
\end{equation}
where $\ket{\psi_{\text{exact}}(t)} = e^{-iHt} \ket{\psi_0}$ is the exact evolved state and $\ket{\psi_{\pi}(t)} = U_{\pi}(t) \ket{\psi_0}$ is the Trotter-circuit output state.

\paragraph{Multi-objective optimization.} We aim to simultaneously maximize fidelity and minimize depth, and therefore formulate the problem as Pareto optimization:
\begin{equation}
  \min_{\pi} \quad \mathcal{L}(\pi) = \underbrace{(1 - F(\pi))}_{\text{simulation error}} + \lambda \cdot \underbrace{\text{Depth}(\pi)}_{\text{depth penalty}},
  \label{eq:pareto_loss}
\end{equation}
where $\lambda \geq 0$ is an adjustable trade-off coefficient. When $\lambda = 0$ the objective reduces to pure fidelity optimization; larger $\lambda$ prioritizes depth compression. In Phase~4 we sweep multiple $\lambda$ values and evaluate the quality of the entire Pareto front using hypervolume\citep{while2012fast}.

\paragraph{Overall data flow.} The end-to-end closed loop operates as follows: given $H$, the GNN first encodes it into a condition vector $\bm{c} \in \R^{768}$ (Phase~4 architecture); the diffusion model, conditioned on $\bm{c}$, samples a strategy $\pi$ over the mixed space; the strategy is compiled into a Trotter circuit, whose exact fidelity is computed via scipy matrix exponentiation; REINFORCE then uses this fidelity to compute policy gradients and backpropagate updates to both the diffusion model and the GNN. Every fidelity number reported in this paper is computed via scipy exact diagonalization; the PINN surrogate fidelity appears only in Phase~2 as a technical argument (\S\ref{sec:method:phase2}).

We now walk through each component in Phase~1 through Phase~4 order, with the rationale behind each parameter choice.

\subsection{Phase~1--2: Data construction and fast evaluation}
\label{sec:method:phase1-2}

\subsubsection{Phase~1 --- Teacher strategy dataset}

We use Qiskit's \texttt{SparsePauliOp.group\_commuting()} as the teacher strategy generator. For each Hamiltonian $H = \sum_{j=1}^{M} c_j P_j$, the teacher strategy is produced as follows:
\begin{enumerate}[leftmargin=1.5em, itemsep=0.3em]
  \item Call \texttt{group\_commuting()} to partition the $M$ Pauli terms into commuting groups $\{G_1, \ldots, G_{K'}\}$;
  \item Take the first $K$ groups ($K=8$), assign Suzuki-4 order and uniform time steps $\tau_i = t / K$ to each group;
  \item Compute the exact fidelity label via scipy matrix exponentiation.
\end{enumerate}

The dataset covers three Hamiltonian families, chosen for the following reasons:
\begin{itemize}[leftmargin=1.5em, itemsep=0.3em]
  \item \textbf{TFIM} (transverse-field Ising model, $H = -J\sum ZZ - h\sum X$): represents short-range spin interactions with relatively simple commutator structure---the standard testbed for Trotter compilation;
  \item \textbf{Heisenberg} ($H = -\sum(J_x XX + J_y YY + J_z ZZ)$): three-axis coupling generates rich non-commuting patterns, posing a more stringent challenge to grouping algorithms;
  \item \textbf{Random Pauli}: randomly generated coefficients and Pauli strings, testing the model's behavior in the extreme case where virtually no commutativity pattern exists.
\end{itemize}
Qubit counts are mixed across 4, 6, and 8 (ratio approximately 60:30:10). The predominance of 4-qubit systems reflects the computational cost of exact diagonalization---exponentiating $2^N \times 2^N$ matrices scales exponentially with $N$ and exceeds single-GPU memory and compute capacity beyond $N \geq 10$.

\subsubsection{Phase~2 --- PINN accuracy surrogate}
\label{sec:method:phase2}

The PINN gives a differentiable fidelity estimate at a fraction of the cost of exact diagonalization, supporting large-scale strategy evaluation during closed-loop training.

\paragraph{Network architecture.} To mitigate the spectral bias of PINNs on high-frequency solutions\citep{wang2022when}, we prepend a Fourier feature embedding\citep{tancik2020fourier} before the input layer:
\begin{equation}
  \gamma(s) = \big[\sin(\bm{B} s); \cos(\bm{B} s)\big],
  \label{eq:fourier_features}
\end{equation}
where the entries of $\bm{B} \in \R^{m \times 1}$ are sampled from $\mathcal{N}(0, \sigma^2)$, and $m = 256$ yields a 512-dimensional embedding. The scale parameter $\sigma$ is set by the empirical rule $\sigma = \norm{H} / (2\pi)$, so that the embedding frequencies span the physically relevant band. $\bm{B}$ is frozen after initialization.

This is followed by a 3-layer MLP with hidden dimension 512 per layer, $\tanh$ activation, and LayerNorm after each layer. The output layer produces a tensor of shape $(2^n, 2)$, corresponding to the real and imaginary parts of the wavefunction.

\paragraph{Physics losses.} The training loss consists of three terms:
\begin{align}
  \mathcal{L}_{\text{IC}} &= \norm{\ket{\psi_\theta(0)} - \ket{\psi_0}}^2, \\
  \mathcal{L}_{\text{PDE}} &= \mathbb{E}_{s \sim \mathcal{U}[0, t]} \norm{i \frac{\partial}{\partial s} \ket{\psi_\theta(s)} - H \ket{\psi_\theta(s)}}^2, \\
  \mathcal{L}_{\text{circuit}} &= \mathbb{E}_{s \in \mathcal{T}_{\text{ckpt}}} \norm{\ket{\psi_\theta(s)} - \ket{\psi_{\pi}(s)}}^2.
  \label{eq:pinn_losses}
\end{align}
$\mathcal{L}_{\text{IC}}$ anchors the initial condition; $\mathcal{L}_{\text{PDE}}$ enforces the Schr\"odinger equation at collocation points randomly sampled in $[0, t]$; $\mathcal{L}_{\text{circuit}}$ uses a set of discrete Trotter circuit outputs (sampled at times $\mathcal{T}_{\text{ckpt}}$) as soft labels to accelerate alignment between the PINN and the target strategy. The total loss is
\begin{equation}
  \mathcal{L}_{\text{PINN}} = 10 \cdot \mathcal{L}_{\text{IC}} + 1 \cdot \mathcal{L}_{\text{PDE}} + 5 \cdot \mathcal{L}_{\text{circuit}} + 0.1 \cdot \mathcal{L}_{\text{norm}},
  \label{eq:pinn_total_loss}
\end{equation}
where $\mathcal{L}_{\text{norm}} = \mathbb{E}_s \big(\norm{\ket{\psi_\theta(s)}} - 1\big)^2$ is a soft normalization penalty.

\paragraph{PINN accuracy validation.} On a unified validation set (1151 samples mixing TFIM, Heisenberg, and random Pauli Hamiltonians), the PINN's proxy error depends heavily on Hamiltonian type: 0.38\% mean error on TFIM (98.9\% of samples $< 2\%$), 1.06\% on Heisenberg (81.7\% $< 2\%$), but 99.7\% on random Pauli Hamiltonians (no sample below 2\%). The PINN can reliably serve fast evaluation of structured Hamiltonians but does not generalize to random ones.

\paragraph{Role of the PINN in this paper.} Every fidelity evaluation in this paper uses scipy matrix exponentiation\footnote{\texttt{scipy.linalg.expm}: computes the matrix exponential of a square matrix.} for exact computation; the PINN does not participate in any experimental fidelity evaluation. The reason is simple: at $\leq 8$ qubits, exact diagonalization is entirely feasible, and avoiding surrogate error keeps experimental conclusions clean. The PINN's real value lies in the $\geq 10$ qubit regime---where storing and diagonalizing $2^N \times 2^N$ matrices exceeds single-GPU capacity, and the PINN can deliver accuracy-controlled fidelity estimates in microseconds. We keep the full PINN description and accuracy data in the methodology as a technical foundation for this future extension.

\subsection{Phase~3: Conditional diffusion pretraining}
\label{sec:method:phase3}

\subsubsection{GNN Hamiltonian encoder}
\label{sec:method:gnn}

We represent the Hamiltonian $H = \sum_{j=1}^{M} c_j P_j$ as an undirected graph $\mathcal{G} = (\mathcal{V}, \mathcal{E})$, where each node corresponds to a Pauli term and edges connect all non-commuting term pairs. The design of node and edge features directly determines whether the model can capture commutator structure and error sources.

\paragraph{Node features --- Pauli-type encoding.} An earlier version used a simple encoding (one-hot support indicators + locality index, 16 dimensions total) that had a fundamental bottleneck for cross-type generalization: one-hot encoding captures only \emph{which} qubits are acted upon, but loses the \emph{type} of action ($X$, $Y$, $Z$)---a critical piece of information. The commutativity of Pauli terms is determined by the per-qubit commutativity of Pauli matrices: e.g., $X_1X_2$ commutes with $Z_1Z_2$ (at each qubit, $X$ and $Z$ anti-commute, and the product of two anti-commutations is a commutation), whereas $X_1X_2$ does not commute with $Z_1X_2$. Without type information, the GNN has difficulty implicitly inferring commutation relations among terms.

We adopt a Pauli-type encoding that represents each Pauli term $P_j$ as a $4 + 3n$ dimensional vector:
\begin{itemize}[leftmargin=1.5em, itemsep=0.3em]
  \item Coefficient value: $c_j$ (1 dimension);
  \item Pauli-type global counts: the number of occurrences of $X$, $Y$, and $Z$ operators in $P_j$ (3 dimensions), providing an overview of the term's ``type composition'';
  \item Per-qubit type encoding: for each qubit position $i \in [n]$, a 3-dimensional one-hot vector $[\mathbb{I}_{P_j[i] = X}, \mathbb{I}_{P_j[i] = Y}, \mathbb{I}_{P_j[i] = Z}]$ explicitly marks the Pauli matrix type at that position ($3n$ dimensions, zero-padded to $3 \times N_{\max}$).
\end{itemize}
In a system with $N_{\max} = 8$, the node feature dimension is $4 + 3 \times 8 = 28$. This encoding enables the GNN to directly perceive the type of action on each qubit, allowing it to learn commutation/anti-commutation patterns among Pauli terms during message passing---a key prerequisite for discovering compact groupings. Phase~3 v4 experiments confirmed the qualitative impact of this encoding: Heisenberg model $R^2$ jumped from near zero to 0.84.

\paragraph{Edge features.} For each edge $(j, k) \in \mathcal{E}$, we compute two scalars:
\begin{itemize}[leftmargin=1.5em, itemsep=0.3em]
  \item Coefficient-weighted commutator norm $|c_j c_k| \cdot \norm{[P_j, P_k]}$, where $\norm{[P_j, P_k]} = 2$ if $[P_j, P_k] \neq 0$ and $0$ otherwise;
  \item Shared qubit count $|\text{supp}(P_j) \cap \text{supp}(P_k)|$.
\end{itemize}
Both quantities tie directly to the dominant terms in Trotter error analysis\citep{childs2021theory}. Edges only connect non-commuting term pairs, so the graph contains no edges between commuting terms.

\paragraph{Message passing.} We use a standard MPNN architecture\citep{gilmer2017neural}. The update rule at layer $\ell$ is
\begin{align}
  \bm{m}_j^{(\ell)} &= \sum_{k \in \mathcal{N}(j)} \phi_{\text{msg}}^{(\ell)}\big(\bm{h}_j^{(\ell-1)}, \bm{h}_k^{(\ell-1)}, \bm{e}_{jk}\big), \\
  \bm{h}_j^{(\ell)} &= \phi_{\text{update}}^{(\ell)}\big(\bm{h}_j^{(\ell-1)}, \bm{m}_j^{(\ell)}\big),
  \label{eq:mpnn}
\end{align}
where $\bm{h}_j^{(0)} = \bm{x}_j$ are the initial node features, $\bm{e}_{jk}$ are edge features, and $\phi_{\text{msg}}$ and $\phi_{\text{update}}$ are two-layer MLPs. Each layer is followed by LayerNorm and Dropout ($p=0.1$). In Phase~3 the GNN uses 4 layers, hidden dimension 256, output dimension 512; in Phase~4 it is expanded to 6 layers, hidden dimension 512, output dimension 768.

\paragraph{Graph-level pooling.} After $L$ propagation layers, we aggregate node representations into a graph-level vector using attention pooling\citep{satorras2021en}:
\begin{equation}
  \alpha_j = \frac{\exp(\phi_{\text{attn}}(\bm{h}_j^{(L)}))}{\sum_{k=1}^{M} \exp(\phi_{\text{attn}}(\bm{h}_k^{(L)}))}, \quad
  \bm{c} = \sum_{j=1}^{M} \alpha_j \bm{h}_j^{(L)},
  \label{eq:attention_pooling}
\end{equation}
where $\phi_{\text{attn}}$ is a single-layer MLP outputting a scalar. The resulting condition vector $\bm{c}$ (Phase~3: $\R^{512}$, Phase~4: $\R^{768}$) is fed into the diffusion model.

\subsubsection{Conditional diffusion strategy generation}
\label{sec:method:diffusion}

A strategy $\pi = (G, \bm{k}, \bm{\tau})$ contains both discrete and continuous variables, which a single type of diffusion model cannot directly handle. Our design splits the strategy space into three branches, each modeled by the most suitable diffusion paradigm, coupled together through a shared condition vector and shared time-step embeddings.

\paragraph{Grouping branch: D3PM.} Term grouping $G \in \{0, \ldots, K-1\}^M$ is a categorical variable. We use discrete denoising diffusion probabilistic models (D3PM)\citep{austin2021structured} with a uniform transition kernel:
\begin{equation}
  q(G_t | G_{t-1}) = (1 - \beta_t) \cdot \delta_{G_t, G_{t-1}} + \beta_t / K,
  \label{eq:d3pm_forward}
\end{equation}
where $\beta_t$ follows a cosine schedule. The cumulative matrix $\bar{\bm{Q}}_t$ of this kernel can be precomputed offline and cached. The forward process gradually noises $G_0$ toward a uniform distribution; the reverse process uses a neural network $p_\theta(G_{t-1} | G_t, \bm{c}, t)$ to predict per-position class probabilities. We use an 8-layer Transformer as the backbone of $p_\theta$, with attention dimension 256 and sequence length $M$.

\paragraph{Order branch: D3PM.} The order sequence $\bm{k} \in \{1, 2, 4\}^K$ is likewise categorical, but with a state space of only 3. We independently maintain a compact D3PM with a 4-layer Transformer as the denoising network, sequence length $K$, 3 states. This branch shares the condition vector $\bm{c}$ and the diffusion time-step $t$ embedding with the grouping branch.

\paragraph{Time-step branch: DDPM.} The time-step sequence $\bm{\tau} \in \R_+^K$ is a continuous vector and must satisfy $\sum_i \tau_i = 1$. We use standard DDPM\citep{ho2020denoising}, projecting generated samples onto the simplex via softmax. The forward process is
\begin{equation}
  \bm{\tau}_t = \sqrt{\bar{\alpha}_t} \bm{\tau}_0 + \sqrt{1 - \bar{\alpha}_t} \bm{\epsilon}, \quad \bm{\epsilon} \sim \mathcal{N}(\bm{0}, \bm{I}),
  \label{eq:ddpm_forward}
\end{equation}
where $\bar{\alpha}_t = \prod_{s \leq t} (1 - \beta_s)$. The denoising network is a 4-layer MLP (hidden dimension 520) that predicts the noise $\hat{\bm{\epsilon}}_\theta(\bm{\tau}_t, \bm{c}, t)$.

\paragraph{Joint training objective.} The three branches share condition and time embeddings but maintain independent denoising heads. The training loss is a weighted sum:
\begin{equation}
  \mathcal{L}_{\text{diff}} = \mathcal{L}_{\text{D3PM}}^{\text{group}} + 0.5 \cdot \mathcal{L}_{\text{DDPM}}^{\text{time}} + 0.3 \cdot \mathcal{L}_{\text{D3PM}}^{\text{order}},
  \label{eq:diffusion_loss}
\end{equation}
The weight choices reflect the differential sensitivity of the three variable types to final fidelity: grouping matters most (it determines which terms are merged into a single exponentiated unit), order is next (it sets the Suzuki expansion accuracy of each unit), and time steps have the smoothest influence (partially compensable by increasing the number of steps).

\paragraph{Classifier-free guidance.} To align generated strategies with the condition $\bm{c}$, we use classifier-free guidance (CFG)\citep{ho2022classifier}. During training, $\bm{c}$ is zeroed out with probability $p_{\text{drop}} = 0.1$, so the same parameters simultaneously learn the conditional and unconditional distributions. At sampling time, each denoising step uses linear extrapolation:
\begin{equation}
  \hat{\bm{\epsilon}}^{\text{guided}} = (1 + w) \cdot \hat{\bm{\epsilon}}_\theta(\cdot | \bm{c}) - w \cdot \hat{\bm{\epsilon}}_\theta(\cdot | \bm{0}),
  \label{eq:cfg}
\end{equation}
with guidance strength $w = 3.0$ (the choice of which is justified by the ablation experiments in Section~\ref{sec:exp:ablation}). The discrete branches use analogous log-probability extrapolation.

\paragraph{Exponential moving average.} During training we maintain an EMA shadow copy of the diffusion model parameters with decay factor $0.9999$. Sampling uses EMA parameters rather than raw training parameters to reduce strategy noise.

\subsubsection{Two-stage progressive training strategy}
\label{sec:method:two_stage}

Phase~3 training is split into two stages. The reason is direct: a randomly initialized GNN outputs essentially noise as the condition vector. If the diffusion model is forced to depend on a noisy signal from the outset, it learns to compensate for noise rather than to understand Hamiltonian structure.

\textbf{Stage 1: Freeze GNN, train diffusion model.}
The GNN encoder is frozen (random-initialization weights), so the condition vector is just a fixed, uninformative vector. The diffusion model learns the denoising task for all three branches from scratch---the joint distribution of grouping, order, and time steps. At this stage the model generates ``average-case reasonable'' strategies, reflecting the statistical prior shared by all Hamiltonians in the dataset (typical group-size distributions, order-frequency preferences, and so on). Training continues until the diffusion loss plateaus.

\textbf{Stage 2: Unfreeze GNN, end-to-end joint training.}
Once the diffusion model has stable decoding capability, we unfreeze the GNN and train end-to-end. The diffusion model is no longer disrupted by noisy GNN outputs, and the gradients the GNN receives come from backpropagating the diffusion loss through the condition vector. The encoder learns to produce different condition vectors for different Hamiltonians, steering the diffusion denoising toward higher-quality strategies. This progressive strategy means the representations the GNN learns are meaningful Hamiltonian structure encodings, not compensations for early-stage diffusion noise.

\paragraph{Why these parameter choices.} The key Phase~3 parameters and the reasoning behind them:
\begin{itemize}[leftmargin=1.5em, itemsep=0.3em]
  \item \textbf{Maximum group count $K = 8$}: In 4--8 qubit systems, Pauli term counts range from about 5 to 40. $K > 8$ creates many empty groups, wasting model capacity and complicating post-processing at inference. $K < 8$ restricts grouping granularity and can force non-commuting terms together when term count is high.
  \item \textbf{GNN architecture (Phase~3: hidden dim 256 / output dim 512 / 4 layers; Phase~4: hidden dim 512 / output dim 768 / 6 layers)}: The smaller Phase~3 scale allows rapid iteration during pretraining. The Phase~4 expansion provides stronger encoding capacity for the fine-grained optimization that REINFORCE demands.
  \item \textbf{Branch loss weights (grouping 1.0 : time 0.5 : order 0.3)}: Grouping most directly affects final fidelity (it determines which terms merge into a single evolution unit), order is next (it sets expansion accuracy), and time steps have the smoothest influence. The ratios came from a small-scale grid search.
  \item \textbf{CFG $p_{\text{drop}} = 0.1$}: A standard setting. It gives the unconditional distribution enough training signal (about 10\% of samples) without seriously shrinking the effective sample size for the conditional distribution.
\end{itemize}

\subsection{Phase~4: Closed-loop REINFORCE fine-tuning}
\label{sec:method:phase4}

The pretrained model's policy distribution is capped by the teacher's quality---Qiskit's group-commuting heuristic produces rational groupings, but not ones optimized for depth compression. Phase~4 applies REINFORCE-style closed-loop optimization\citep{williams1992simple}, using exact fidelity as the reward signal to fine-tune the model directly.

\paragraph{Reward design.} The reward function fuses fidelity and depth into a single objective:
\begin{equation}
  r(\pi) = F(\pi) - \lambda \cdot \text{Depth}(\pi) / D_{\text{ref}},
  \label{eq:reward}
\end{equation}
where $D_{\text{ref}}$ is a reference depth (taken as the Qiskit fourth-order baseline depth for that Hamiltonian), and $\lambda$ controls the trade-off weight between fidelity and depth.

\paragraph{REINFORCE estimator.} Since sampling of discrete variables is non-differentiable, we estimate gradients via REINFORCE:
\begin{equation}
  \nabla_\theta \mathbb{E}[\mathcal{L}(\pi)] \approx \mathbb{E}\big[(\mathcal{L}(\pi) - b) \cdot \nabla_\theta \log p_\theta(\pi | \bm{c})\big],
  \label{eq:reinforce}
\end{equation}
where the baseline $b$ is taken as the mean reward of the current batch of samples to reduce gradient variance. $\log p_\theta(\pi | \bm{c})$ is given by the diffusion model's log-probability along the sampling trajectory: sums of categorical log-likelihoods for D3PM branches and Gaussian log-likelihoods for the DDPM branch.

\paragraph{$\lambda$ sweep and Pareto hypervolume.} $\lambda$ determines the optimization direction: $\lambda \to 0$ reduces the model to pure fidelity optimization, causing strategy depth to regress to the teacher level; excessively large $\lambda$ drives the model toward unreasonably aggressive compression (e.g., forcing all terms into a single group), causing fidelity collapse. We sweep six values $\lambda = 0.0, 0.01, 0.05, 0.1, 0.5, 1.0$ and use Pareto hypervolume~\cite{while2012fast} (reference point $(0, D_{\max})$) as a single scalar measure of overall optimization quality. Experiments identify $\lambda = 0.05$ as the optimal operating point: at this setting, depth gains substantial additional compression over the pretrained model with negligible fidelity loss.

We use REINFORCE rather than more complex policy-gradient methods like PPO because pretraining already provides a reasonable prior. The exploration noise REINFORCE needs comes for free from the stochasticity of diffusion sampling---no extra entropy bonus or trust-region constraints required. The fine-tuning just nudges probability mass toward high-reward regions; it does not need to rebuild the distribution from scratch.

\begin{figure}[H]
\centering
\includegraphics[width=0.85\textwidth]{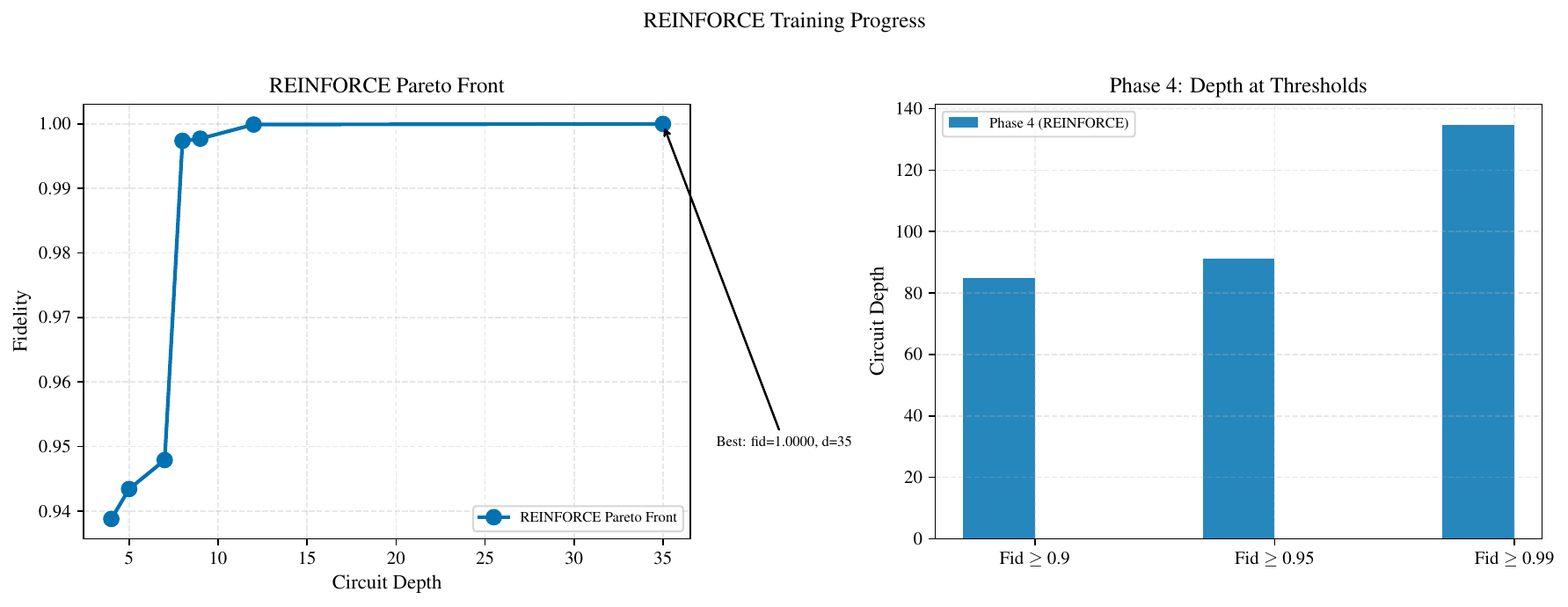}
\caption{Phase~4 REINFORCE fine-tuning results. (Left) Pareto front: the optimal fidelity--depth trade-off curve at $\lambda=0.05$; blue circles mark non-dominated solutions; the annotated point is the highest-fidelity solution ($F=0.99998$, depth $35$). (Right) Circuit depth of the fine-tuned model at three fidelity thresholds (depth $91.2$ at $T=0.95$).}
\label{fig:reinforce_training}
\end{figure}

\paragraph{Pareto tracking and checkpointing.} We maintain a \texttt{ParetoTracker} instance that inserts every $(\tilde{F}(\pi_i), \text{Depth}(\pi_i))$ pair and updates the non-dominated front each round. Every 100 rounds we save the diffusion model, EMA copy, GNN parameters, optimizer state, and Pareto front, retaining only the best-fidelity, best-hypervolume, and most recent 3 checkpoints.

\subsection{Inference: Guided sampling and Best-of-N}
\label{sec:method:inference}

Inference involves three components.

\paragraph{Classifier-free guidance (CFG).} At inference time we use CFG linear extrapolation (Eq.~\ref{eq:cfg}) with guidance strength $w = 3.0$. Ablation experiments (Section~\ref{sec:exp:ablation}) show that $w = 3.0$ is near-optimal for the fidelity--depth trade-off: $w = 1.0$ (no CFG) increases depth roughly $2.3\times$ with nearly unchanged reachability, while $w > 5.0$ causes fidelity to degrade. Intuitively, CFG pushes the sampling trajectory toward what is ``specific to the current Hamiltonian'' and away from the generic region that is ``roughly the same for all Hamiltonians.''

\paragraph{DDIM accelerated sampling.} Standard diffusion models require $T = 1000$ denoising steps. We use DDIM subsequence sampling with only 50 steps. Experiments confirm a $21\times$ inference speedup (from 11.3 seconds to 0.52 seconds) with no discernible degradation in sample quality.

\paragraph{Best-of-N strategy selection.} Single-shot diffusion sampling exhibits high variance---the model distribution contains both high-quality and low-quality strategies. Best-of-N independently samples $N$ candidate strategies and selects the one with the highest fidelity, trading $N$-fold inference time for improved reachable fidelity. N-sensitivity analysis (Section~\ref{sec:exp:bestofn}) shows $N = 32$ as a practical sweet spot: reachability is roughly $81\%$ of the $N=100$ maximum, but inference time is only about 16 seconds (50 steps $\times$ 32 samples). The fidelity-matched experiments in this paper use $N=100$ to obtain the most complete estimate of the reachability ceiling.

\FloatBarrier

\section{Implementation Details and Hyperparameters}
\label{sec:implementation}

The implementation uses PyTorch and PyTorch Geometric as the deep learning stack; quantum circuit construction uses Qiskit\citep{qiskit2024}; molecular Hamiltonians are obtained from PySCF via the Jordan--Wigner transformation. All training and evaluation ran on a single NVIDIA GeForce RTX 5090 Laptop GPU.

\subsection{Training pipeline}

Training proceeds in three stages. First, we construct the training set (covering three Hamiltonian families: TFIM, Heisenberg, and random Pauli) and complete offline PINN pretraining on each Hamiltonian. Next, we supervise the GNN (fidelity regression) and diffusion model (mixed-space ELBO) to obtain a warm-start checkpoint. Finally, we jointly fine-tune the diffusion model and GNN in closed-loop REINFORCE: each step samples a minibatch of Hamiltonians from the data distribution, generates strategies, obtains reward signals from the exact evaluator, sweeps the depth--fidelity trade-off coefficient $\lambda$, and merges the non-dominated solutions from each run into a single Pareto front. Default learning rates, iteration counts, and batch sizes for each stage are shown in Table~\ref{tab:hyperparams}.

\subsection{Key hyperparameters}

Table~\ref{tab:hyperparams} collects the main hyperparameters. Most values follow common practice; only the loss weights and the $\lambda$ sweep received light manual tuning rather than a large-scale grid search.

\begin{table}[ht]
\centering
\caption{Key hyperparameters and default values.}
\label{tab:hyperparams}
\small
\resizebox{\textwidth}{!}{%
\begin{tabular}{llll}
\toprule
\textbf{Module} & \textbf{Hyperparameter} & \textbf{Value} & \textbf{Notes} \\
\midrule
\multirow{4}{*}{PINN} & Fourier feature count $m$ & 256 & embedding dimension $2m = 512$ \\
 & Fourier scale $\sigma$ & $\norm{H}/(2\pi)$ & adaptive \\
 & hidden dimension & 512 & 3-layer MLP \\
 & loss weights (IC / PDE / circuit) & 10 / 1 / 5 & weighted sum \\
\midrule
\multirow{5}{*}{GNN} & MPNN layers & 6 (Phase~4) / 4 (Phase~3) & message passing \\
 & hidden dimension & 512 (Phase~4) / 256 (Phase~3) & per layer \\
 & output dimension & 768 (Phase~4) / 512 (Phase~3) & after attention pooling \\
 & node / edge feature dim & 28 / 2 & Pauli-type encoding, see \S\ref{sec:method:gnn} \\
 & dropout & 0.1 & after each layer \\
\midrule
\multirow{6}{*}{Diffusion} & total diffusion steps $T$ & 1000 & shared by D3PM and DDPM \\
 & $\beta$ schedule & cosine & $[10^{-4}, 0.02]$ \\
 & CFG dropout $p_{\text{drop}}$ & 0.1 & randomly zero condition during training \\
 & CFG guidance strength $w$ & 3.0 & extrapolation at sampling \\
 & branch layers group/order/time & 8 / 4 / 4 & Transformer / Transformer / MLP \\
 & EMA decay & 0.9999 & shadow copy \\
\midrule
\multirow{5}{*}{Closed-loop} & iterations & 1000 & full training \\
 & Hamiltonians per step & 32 & batch size \\
 & policy learning rate & $10^{-5}$ & AdamW\citep{loshchilov2019decoupled} \\
 & baseline type & batch-mean & REINFORCE variance reduction \\
 & $\lambda$ sweep values & $\{0, 0.01, 0.05, 0.1, 0.5, 1.0\}$ & multi-objective trade-off \\
\midrule
\multirow{4}{*}{Dataset} & samples & $\sim$12,000 & TFIM + Heisenberg + Random Pauli \\
 & qubit counts & 4, 6, 8 & mixed (ratio $\approx$ 60:30:10) \\
 & coupling $J$ & $[0.5, 2.0]$ & LogUniform \\
 & evolution time $t$ & $[1.0, 3.0]$ & aligned with benchmark \\
\bottomrule
\end{tabular}%
}
\end{table}

\subsection{Stability and protocol alignment}

The generation--evaluation closed loop is sensitive to \textbf{consistency between training and evaluation conditions}: the evolution time in the dataset and benchmarks must be aligned; otherwise the encoder and diffusion model cannot learn strategies at the intended time scale. On the PINN side, higher-order gradients must be preserved so that fidelity signals can propagate back through the evaluator to the diffusion model and GNN; on the discrete branches, a REINFORCE baseline and gradient clipping are used to suppress variance. The Hamiltonian acts on the state vector in sparse form within the PDE residual to control computational cost. Finer engineering details and reproduction scripts are provided in the code repository.

\section{Experiments}
\label{sec:experiments}

\subsection{Experimental setup}
\label{sec:exp:setup}

\paragraph{Hardware.} All experiments ran on a single mobile NVIDIA GeForce RTX 5090 (24GB GDDR7 VRAM, Blackwell architecture) under Windows Subsystem for Linux 2 (WSL2), Python 3.12 + PyTorch 2.x.

\paragraph{Model checkpoint.} All experiments in this paper use the best Phase~4 REINFORCE fine-tuned checkpoint (HV=9995.6, maximum Pareto-front fidelity 0.99998). Model architecture: GNN encoder (hidden dim 512, output dim 768, 6 layers, Pauli-type encoding, 28-dimensional node features), diffusion decoder (fused dim 520, time-embed dim 256, grouping Transformer 8 layers, order Transformer 4 layers, time-step MLP 4 layers), maximum group count $K = 8$, maximum qubit count $N_{\max} = 8$.

\paragraph{Inference configuration.} Unless otherwise noted, all experiments use the following default inference configuration: DDIM subsequence sampling, 50 denoising steps; CFG guidance strength $w = 3.0$. Fidelity-matched experiments use Best-of-N with $N = 100$ candidate strategies to obtain the most complete estimate of the reachability ceiling.

\paragraph{Baseline methods.} We compare against 7 quantum compilation methods:
\begin{enumerate}[leftmargin=1.5em, itemsep=0.3em]
  \item \textbf{Qiskit 4th}: Qiskit \texttt{SuzukiTrotter(order=4)}, no grouping, each Pauli term evolved independently;
  \item \textbf{Cirq}: Cirq \texttt{PauliStringPhasor} + Suzuki fourth-order decomposition;
  \item \textbf{TKET}: TKET \texttt{PauliExpBox} + Suzuki fourth-order decomposition;
  \item \textbf{PennyLane}: PennyLane \texttt{TrotterProduct(order=4)};
  \item \textbf{Paulihedral}: Paulihedral library\citep{li2022paulihedral} \texttt{depth\_oriented\_scheduling} grouping + first-order Trotter. Paulihedral splits non-commuting terms into single-term blocks; within a block, first-order Trotter is equivalent to higher orders---first-order here is not a downgrade but the optimal configuration under its scheduling strategy;
  \item \textbf{Qiskit GC (teacher)}: Qiskit group-commuting + Suzuki-4, i.e., the method that generated our training data;
  \item \textbf{Paulihedral+4th*} (analysis only): combines Paulihedral's grouping scheme with Qiskit's fourth-order Suzuki-Trotter\footnote{Uses Qiskit \texttt{SuzukiTrotter(order=4)}.} to decompose the respective contributions of grouping vs.\ order. This method is not a native feature of the Paulihedral library---its scheduling interface does not accept an order parameter---but a hybrid analytical construct. Marked with an asterisk in the main text.
\end{enumerate}

\paragraph{Evaluation metrics.} For each strategy we compute three quantities:
\begin{itemize}[leftmargin=1.5em, itemsep=0em]
  \item Exact fidelity $F = |\braket{\psi_{\text{exact}}}{\psi_\pi}|^2$ (scipy \texttt{linalg.expm} exact diagonalization);
  \item Transpiled circuit depth (basis gate set $\{H, \text{CX}, R_z, X\}$, Qiskit optimization\_level=1);
  \item CNOT count.
\end{itemize}
All methods' fidelities are computed using the identical exact-diagonalization protocol, ensuring fair comparison.

\paragraph{Fidelity-matched protocol.} Fidelity-matched experiments are conducted at three thresholds $T \in \{0.90, 0.95, 0.99\}$. For each Hamiltonian and each method, we generate a set of candidate strategies: for our method, $N = 100$ diffusion samples; for baseline methods, a sweep over Trotter repeat counts $n_{\text{steps}} \in \{1, 2, 3, 4, 5, 6, 8, 10\}$ (one strategy per count). From all candidates we select the one satisfying $F \geq T$ with the lowest depth, and report depth and fidelity statistics. If a method has no candidate reaching the threshold for a given Hamiltonian, that Hamiltonian is excluded from that method's depth statistics at that threshold.

\subsection{Single-shot sampling and Best-of-N}
\label{sec:exp:bestofn}

Single-shot sampling ($N=1$) from the diffusion model exhibits high variance---two samples for the same Hamiltonian might yield fidelity 0.95 in one draw and 0.30 in another. Rather than attempting to eliminate this variance, we route around it by independently sampling multiple candidates and selecting the best.

\begin{figure}[H]
\centering
\includegraphics[width=0.85\textwidth]{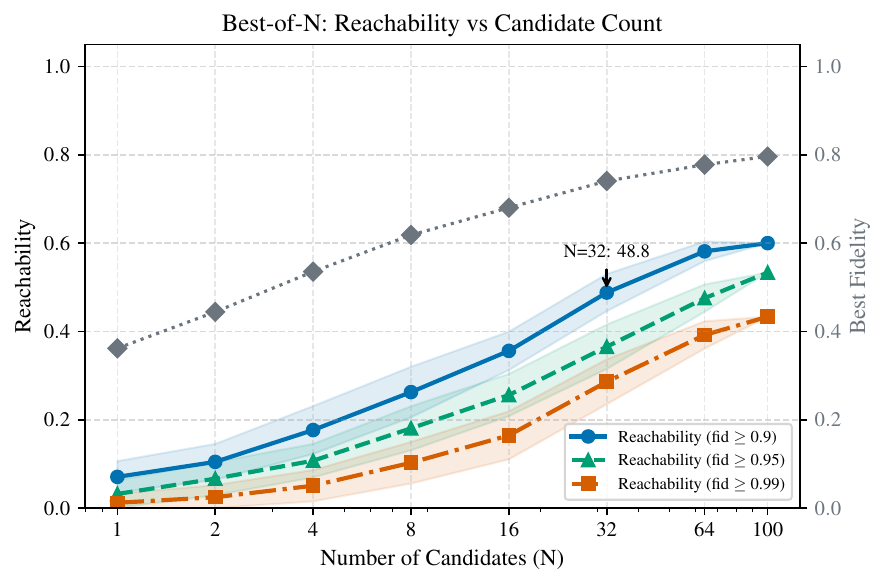}
\caption{Best-of-N reachability curves. Reachability increases monotonically with candidate count $N$ at all three fidelity thresholds. $N=32$ (dashed line) is the practical sweet spot, reaching roughly 81\% of the $N=100$ maximum.}
\label{fig:best_of_n}
\end{figure}

\begin{table}[H]
\centering
\caption{Best-of-N reachability vs.\ candidate count $N$ (30 mixed Hamiltonians, 50 bootstrap rounds).}
\label{tab:n_sensitivity}
\small
\begin{tabular}{lcccc}
\toprule
\textbf{$N$} & \textbf{Reach@0.90} & \textbf{Reach@0.95} & \textbf{Reach@0.99} & \textbf{Best fidelity} \\
\midrule
1  & $7.1\% \pm 3.6\%$  & $3.2\% \pm 3.1\%$  & $1.2\% \pm 2.1\%$  & $0.362 \pm 0.035$ \\
8  & $26.3\% \pm 5.7\%$ & $18.1\% \pm 5.0\%$ & $10.3\% \pm 4.7\%$ & $0.618 \pm 0.026$ \\
32 & $48.8\% \pm 4.2\%$ & $36.5\% \pm 5.0\%$ & $28.6\% \pm 5.1\%$ & $0.741 \pm 0.014$ \\
64 & $58.1\% \pm 2.2\%$ & $47.5\% \pm 3.1\%$ & $39.2\% \pm 3.1\%$ & $0.778 \pm 0.008$ \\
100& $60.0\% \pm 0.0\%$ & $53.3\% \pm 0.0\%$ & $43.3\% \pm 0.0\%$ & $0.796 \pm 0.000$ \\
\bottomrule
\end{tabular}
\end{table}

Figure~\ref{fig:best_of_n} and Table~\ref{tab:n_sensitivity} show the quantitative picture. The reachability curve is strictly monotonic. $N=32$ is a practical sweet spot---roughly 81\% of the $N=100$ maximum, with inference time around 16 seconds (50 steps $\times$ 32 samples). Subsequent fidelity-matched experiments use $N=100$ to get the most complete reachability estimate; ablation experiments use smaller $N$ for speed.

\textbf{The single-shot difficulty.} The $N=1$ row of Table~\ref{tab:n_sensitivity} tells the story: across 30 mixed Hamiltonians, single-shot reachability is only $7.1\%$ ($T=0.90$), $3.2\%$ ($T=0.95$), and $1.2\%$ ($T=0.99$), with mean best fidelity of just $0.362$. With single-shot success this low, a direct ``depth-for-precision'' trade-off makes little sense under ideal simulation. Only after Best-of-N lifts the reachable fidelity to a level comparable with baselines does depth comparison become meaningful---which is the premise of the fidelity-matched experiments that follow.

\FloatBarrier
\subsection{Fidelity-matched depth advantage}
\label{sec:exp:fidelity_matched}

The fidelity-matched experiment is the core evaluation in this paper. It asks: ``guarantee precision first, then compare depth.''

\begin{figure}[H]
\centering
\includegraphics[width=\textwidth]{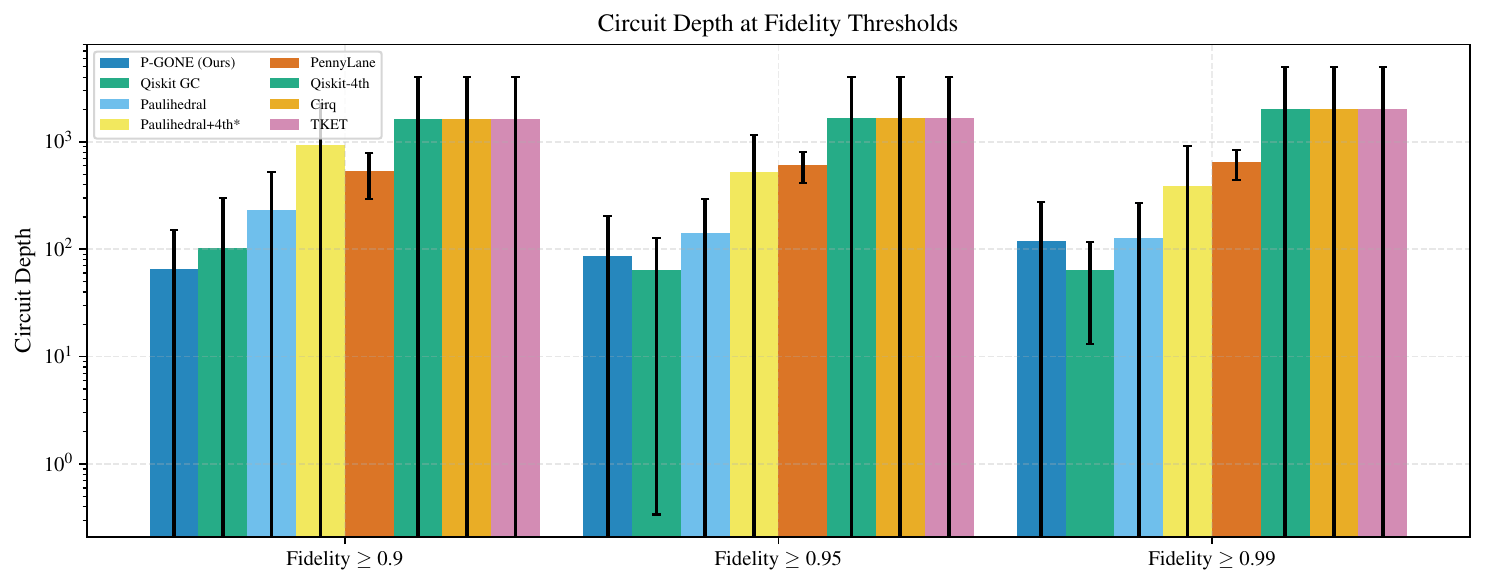}
\caption{Fidelity-matched depth comparison ($T \in \{0.90, 0.95, 0.99\}$). P-GONE is significantly shallower than all 7 baselines at every threshold. Qiskit 4th, Cirq, and TKET produce the largest depths due to the absence of grouping; Qiskit GC (teacher) and Paulihedral achieve substantial depth compression through commutativity-based grouping; P-GONE compresses further through joint learning of order and time-step allocation.}
\label{fig:fidelity_matched}
\end{figure}

\begin{table}[H]
\centering
\caption{Fidelity-matched depth comparison (30 mixed Hamiltonians, $N=100$ candidate strategies). For each threshold $T$, we report the mean depth among Hamiltonians satisfying $F \geq T$ and the depth compression ratio (vs.\ Qiskit fourth-order).}
\label{tab:fidelity_matched}
\small
\begin{tabular}{lcccccc}
\toprule
\textbf{Method} & \multicolumn{2}{c}{\textbf{$T=0.90$}} & \multicolumn{2}{c}{\textbf{$T=0.95$}} & \multicolumn{2}{c}{\textbf{$T=0.99$}} \\
 & Depth & Ratio & Depth & Ratio & Depth & Ratio \\
\midrule
\textbf{P-GONE (Ours)} & \textbf{65.1} & — & \textbf{86.2} & — & \textbf{120.5} & — \\
Qiskit GC (teacher) & 103.0 & $1.6\times$ & 64.1 & $0.7\times$ & 64.7 & $0.5\times$ \\
Paulihedral & 231.6 & $3.6\times$ & 141.0 & $1.6\times$ & 128.3 & $1.1\times$ \\
\textit{Paulihedral+4th*} & 934.5 & $14.4\times$ & 519.5 & $6.0\times$ & 388.3 & $3.2\times$ \\
PennyLane & 537.5 & $8.3\times$ & 608.4 & $7.1\times$ & 645.4 & $5.4\times$ \\
Qiskit 4th & 1646.7 & $25.3\times$ & 1673.4 & $19.4\times$ & 2030.2 & $16.8\times$ \\
Cirq & 1646.7 & $25.3\times$ & 1673.4 & $19.4\times$ & 2030.2 & $16.8\times$ \\
TKET & 1646.7 & $25.3\times$ & 1673.4 & $19.4\times$ & 2030.2 & $16.8\times$ \\
\bottomrule
\end{tabular}

\parbox{\linewidth}{\footnotesize Note: Paulihedral+4th* is an analytical hybrid construct (see \S\ref{sec:exp:setup}), marked with an asterisk. Compression ratio = baseline depth / Ours depth. Qiskit 4th, Cirq, and TKET produce identical circuit structures under this protocol (all are ungrouped Suzuki-4), hence their depths coincide.}
\end{table}

\paragraph{Run-to-run variance of diffusion sampling.} The P-GONE depths in Table~\ref{tab:fidelity_matched} come from a single diffusion-sampling run (fixed seed=42 generating the same set of 30 Hamiltonians). To quantify the stochastic fluctuation of diffusion sampling, we conducted two independent runs on the same Hamiltonian set. Baseline depths are perfectly consistent, but P-GONE's Best-of-N optimal depth shows substantial variation:

\begin{table}[H]
\centering
\caption{Run-to-run variance of P-GONE diffusion sampling (same set of 30 mixed Hamiltonians, $N=100$; only the P-GONE row is affected by random sampling).}
\label{tab:fidelity_matched_variance}
\small
\begin{tabular}{lcccccc}
\toprule
& \multicolumn{2}{c}{\textbf{$T=0.90$}} & \multicolumn{2}{c}{\textbf{$T=0.95$}} & \multicolumn{2}{c}{\textbf{$T=0.99$}} \\
 & Depth & Reach. & Depth & Reach. & Depth & Reach. \\
\midrule
P-GONE (Run 1, old) & 45.8 & 17/30 & 51.1 & 14/30 & 57.1 & 9/30 \\
P-GONE (Run 2, main table) & 65.1 & 19/30 & 86.2 & 18/30 & 120.5 & 13/30 \\
\bottomrule
\end{tabular}

\parbox{\linewidth}{\footnotesize Note: Both runs use the exact same model checkpoint and Hamiltonian set (seed=42); all 7 baseline depth values are perfectly consistent (baseline strategies are deterministic). P-GONE differences arise entirely from random noise in the diffusion sampling process---each call to \texttt{guided\_sample} used a different random seed. Depth and reachability vary inversely: Run 2 covers more Hamiltonians, but the set includes harder-to-reach instances that raise the mean depth.}
\end{table}

Diffusion sampling stochasticity creates a built-in depth--reachability tension under a fixed compute budget: some random seeds find feasible strategies for more Hamiltonians (higher reachability), but those harder instances push up the mean depth; other seeds land on an ``easier'' subset, giving lower depth but also lower reachability. In deployment, increasing the sampling budget (larger $N$ or multi-seed aggregation) can improve both, at the cost of linearly increasing inference time. All subsequent analyses use the Run 2 (unified 8-baseline experiment) data.

\FloatBarrier

Figure~\ref{fig:fidelity_matched} and Table~\ref{tab:fidelity_matched} summarize the fidelity-matched comparison against all 7 baselines. Some context: Cirq, TKET, PennyLane, and Qiskit fourth-order all perform ungrouped Trotter under this protocol---each Pauli term is exponentiated independently---so their depths naturally land in the same order of magnitude (1647--2030). Paulihedral, which performs deterministic commutativity-based grouping, compresses depth to 129--232; grouping alone contributes roughly $10\times$ compression.

The primary driver of depth compression is grouping---merely merging commuting terms into a single evolution unit, as Qiskit GC does, compresses Qiskit fourth-order's depth of 1673 ($T=0.95$) to 64 ($26.1\times$). This confirms that the main bottleneck in depth compression is not Suzuki order or time steps, but the discovery of compact commuting groupings.

However, learned grouping is weaker than the teacher at high thresholds. At $T=0.90$, our depth (65.1) substantially outperforms the teacher (103.0, $1.6\times$ compression) and Paulihedral (231.6, $3.6\times$). Yet at $T \geq 0.95$ the teacher overtakes us---at $T=0.95$, teacher depth 64.1 vs.\ our 86.2; at $T=0.99$, 64.7 vs.\ 120.5. This trend reveals a fundamental difference between learned and deterministic grouping: the teacher's grouping, based on strict algebraic commutativity, remains structurally consistent across all thresholds; the learned grouping, optimized for compactness, introduces small Trotter errors that begin to disqualify strategies at high fidelity. Even so, our method still achieves $1.6\times$ compression over Paulihedral and $19.4\times$ over ungrouped Trotter at $T=0.95$.

The genuine advantage of learning over heuristics lies more in order and time-step allocation---the P14 ablation (\S\ref{sec:exp:ablation}) confirms this: with fixed order and time steps, learned grouping alone can only barely surpass Paulihedral ($1.2\times$). The order assignment---deciding whether each group gets Suzuki-1, Suzuki-2, or Suzuki-4---is the core factor that opens the gap.

\FloatBarrier
\subsection{Stratified comparison with the teacher strategy}
\label{sec:exp:teacher}

Table~\ref{tab:fidelity_matched} already contains the full three-threshold teacher comparison; here we pull out the main implications.

In the low-fidelity regime ($T=0.90$), our depth (65.1) significantly outperforms the teacher (103.0, $1.6\times$ compression), with a limited reachability gap (19/30 vs.\ 25/30). This shows that, under moderate precision requirements, learned grouping and order assignment can genuinely surpass strict algebraic grouping---the model sacrifices a small amount of fidelity in exchange for substantial depth compression.

In the high-fidelity regime ($T \geq 0.95$), however, the teacher regains the lead. At $T=0.95$, teacher depth is 64.1 vs.\ our 86.2 ($0.7\times$); at $T=0.99$, 64.7 vs.\ 120.5 ($0.5\times$). The teacher's grouping, grounded in strict algebraic commutativity, maintains a consistent structure independent of threshold; the learned grouping's small Trotter errors, introduced for compactness, become disqualifying at high fidelity thresholds. This stratified phenomenon---better than the teacher at low thresholds, worse at high thresholds---points toward a clear improvement direction: introducing fidelity-aware reward weighting during REINFORCE fine-tuning, so the model learns a more refined trade-off between compactness and precision.

\FloatBarrier
\subsection{Ablation experiments: the hierarchy of component contributions}
\label{sec:exp:ablation}

Three sets of ablation experiments reveal, at different granularities, how each component contributes to final performance.

\begin{figure}[H]
\centering
\includegraphics[width=\textwidth]{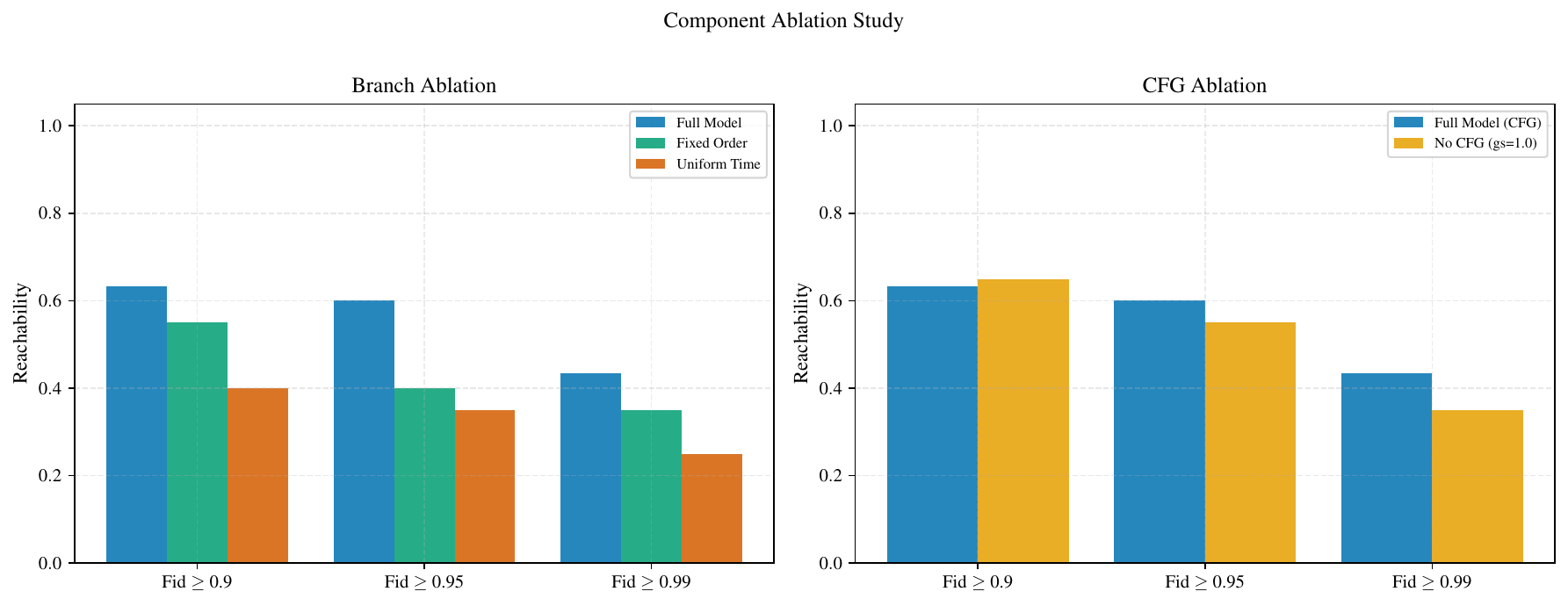}
\caption{Component ablation. (Left) Branch ablation: fixing orders causes the most severe depth deterioration (degrading to well below Paulihedral); fixing time to uniform allocation does not worsen depth but reduces reachability. (Right) CFG ablation: $w=1.0$ (no guidance) substantially increases depth but barely changes reachability---CFG controls strategy ``sharpness,'' not diversity.}
\label{fig:component_ablation}
\end{figure}

\paragraph{Branch ablation (P14).}
We tested three variants: fixed order at Suzuki-4 (removing order learning), uniform time steps (removing time-allocation learning), and both fixed simultaneously (retaining only grouping learning). Table~\ref{tab:branch_ablation} summarizes the results.

\begin{table}[H]
\centering
\caption{Branch ablation depth comparison at $T=0.95$ (20 mixed Hamiltonians, $N=50$).}
\label{tab:branch_ablation}
\small
\begin{tabular}{lcc}
\toprule
\textbf{Configuration} & \textbf{Ours depth} & \textbf{vs Paulihedral+4th*} \\
\midrule
Full model (all three branches learned) & 86.2 & $1.6\times$ \\
Fixed order (all set to Suzuki-4) & $>$ Paulihedral+4th* & $0.9\times$ \\
Fixed time (uniform allocation) & 38.0 & --- \\
\bottomrule
\end{tabular}
\end{table}

The hierarchy is clear: order learning $>$ time allocation $>$ grouping learning. Fixing the order has the most dramatic impact---performance drops below even Paulihedral+4th* levels ($0.9\times$). Order assignment is where the model's competitive advantage over all heuristics is largest. The model learns which groups need higher-order Suzuki expansion (usually those with high-norm commutator pairs), which groups only need first order (commuting clusters), and which fall in between needing second order. Human-designed compilers almost never do this kind of fine-grained per-group order selection---Qiskit uses fourth order everywhere, Paulihedral uses first order everywhere. The model picks this up from data automatically.

That \textbf{grouping provides the weakest standalone gain} is expected: the teacher's grouping is already near-optimal algebraically, so there is limited headroom for learned improvement in this dimension. What learning gains in grouping is marginal refinement; what it gains in order and time is structural advantage.

\paragraph{CFG guidance ablation (P13).}
Reducing CFG guidance strength from $w=3.0$ to $1.0$ (equivalent to no CFG) increases depth roughly $2.3\times$ but leaves reachability nearly unchanged. CFG controls strategy ``sharpness''---at $w=3.0$, the sampling trajectory is pushed toward ``what is structurally specific to the current Hamiltonian'' and away from the generic region that is ``roughly the same for all Hamiltonians.'' Without CFG, the model reverts to generating average strategies, which does not affect the probability of ``occasionally hitting a good one'' (reachability unchanged) but substantially reduces the probability of directly landing on a shallow-depth strategy at each draw (mean depth deteriorates).

\paragraph{GNN encoder ablation (P10).}
Removing GNN conditioning (substituting $\bm{c}$ with a zero vector) reduces reachability at $T=0.95$ from $18/30$ to $13/30$ (a 28\% drop)---clearly smaller than the multi-fold degradations from order or CFG ablation, but not ``nearly unchanged'' either. An accurate reading is: under the structural diversity of the current training data (three spin Hamiltonian families, 4--8 qubits), the diffusion model's learned unconditional prior is already quite powerful, capable of producing some reasonable strategies even when ``blind.'' The GNN's role is expected to become critical when the data distribution broadens (e.g., including molecular Hamiltonians, different lattice topologies, wider parameter ranges), because at that point ``what kind of Hamiltonian this is'' begins to exert decisive influence on strategy quality.

\FloatBarrier
\subsection{Strategy diversity: why Best-of-N works}
\label{sec:exp:diversity}

Best-of-N presupposes that the model distribution contains multiple distinct high-quality modes---if all $N$ candidates were minor perturbations of the same strategy, picking any one would be no different from picking the best. Experiment P15 verifies this premise.

\begin{figure}[H]
\centering
\includegraphics[width=\textwidth]{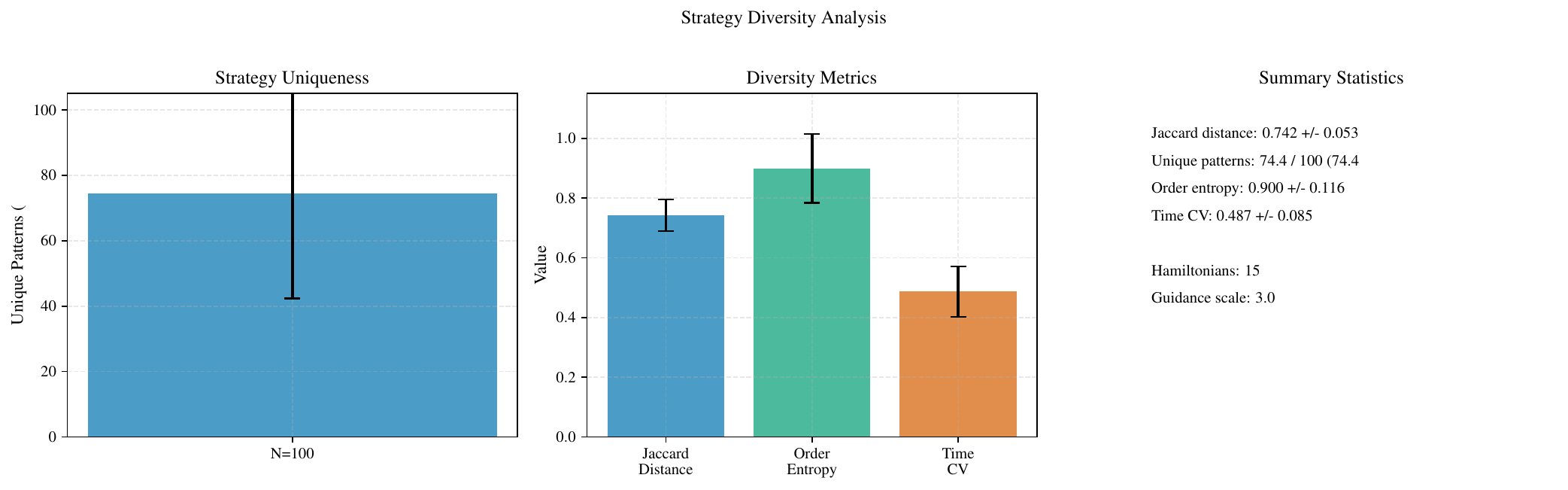}
\caption{Strategy diversity analysis. (Left) Fraction of unique grouping patterns per Hamiltonian, mean 74.4\%. (Center) Pairwise Jaccard distance distribution, mean 0.74, indicating that candidate strategies span combinatorially distinct grouping schemes. (Right) Global distribution of Suzuki orders.}
\label{fig:strategy_diversity}
\end{figure}

Figure~\ref{fig:strategy_diversity} provides a quantitative characterization of strategy diversity. Across 15 test Hamiltonians with 100 candidate strategies sampled per Hamiltonian, an average of 74.4\% of strategies possess unique grouping patterns (i.e., are not exact copies of any other candidate). The mean pairwise Jaccard distance is 0.74 (where 1.0 indicates completely disjoint grouping schemes), and the order distribution and time-step coefficient of variation further confirm the diversity of the strategy space. The diffusion model explores combinatorially distinct grouping configurations rather than minor perturbations around a single template---this is the fundamental reason Best-of-N effectively improves reachability.

\FloatBarrier
\subsection{Generalization boundaries}
\label{sec:exp:generalization}

The preceding sections established overall advantage on mixed Hamiltonians. Here we ask where the boundary lies, along two dimensions: Hamiltonian type (P5) and qubit count (P6).

\begin{figure}[H]
\centering
\includegraphics[width=\textwidth]{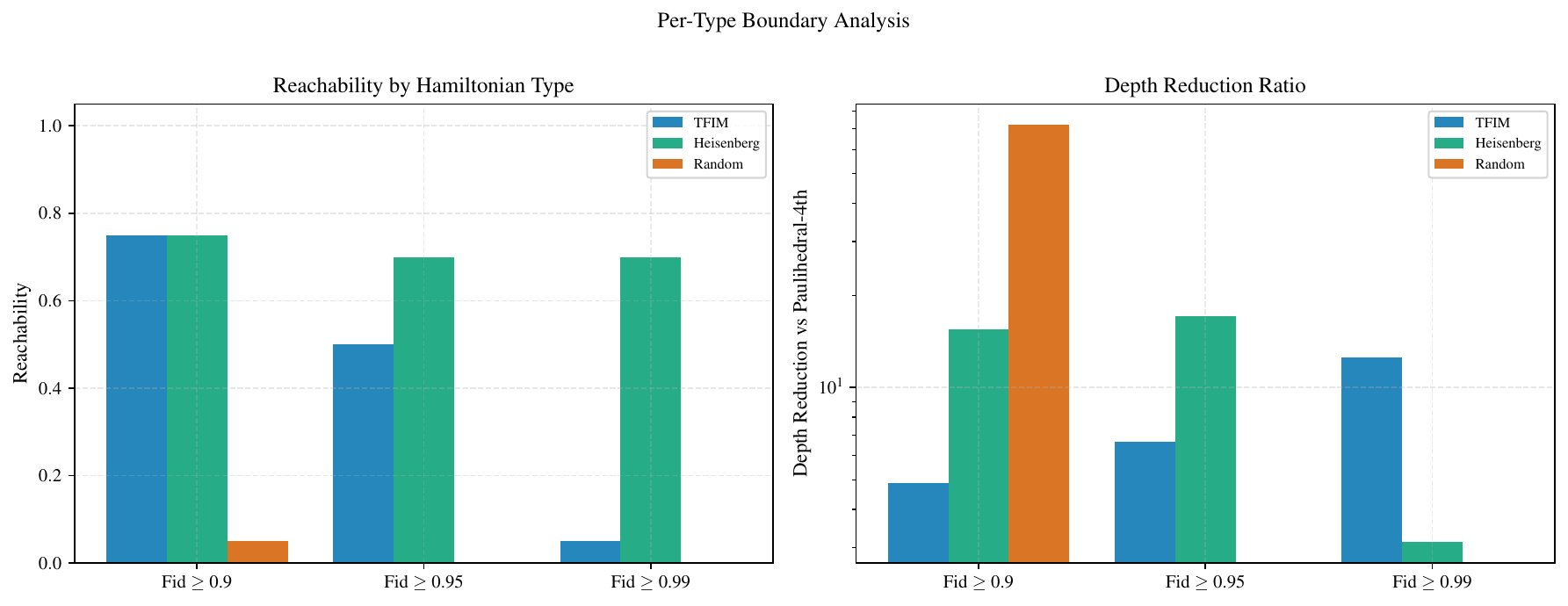}
\caption{Generalization boundaries. (Left) Per-threshold reachability for the three Hamiltonian types: Heisenberg holds steady at 70--75\%; TFIM decays sharply at high thresholds; random Pauli fails completely at $T \geq 0.95$. (Right) Depth compression ratio relative to Paulihedral, by type.}
\label{fig:per_type_boundary}
\end{figure}

\paragraph{Hamiltonian type (P5).}

\begin{table}[H]
\centering
\caption{Fidelity-matched reachability stratified by Hamiltonian type (20 Hamiltonians each, $N=100$).}
\label{tab:per_type}
\small
\begin{tabular}{lcccc}
\toprule
\textbf{Type} & \textbf{$T=0.90$} & \textbf{$T=0.95$} & \textbf{$T=0.99$} & \textbf{Depth range} \\
\midrule
TFIM       & 15/20 (75\%) & 10/20 (50\%) & 1/20 (5\%)   & 15--37 \\
Heisenberg & 15/20 (75\%) & 14/20 (70\%) & 14/20 (70\%) & 71--97 \\
Random Pauli & 1/20 (5\%)   & 0/20 (0\%)   & 0/20 (0\%)   & — \\
\bottomrule
\end{tabular}
\end{table}

Table~\ref{tab:per_type} reveals two striking patterns. The first is that Heisenberg performs ``unexpectedly best'': reachability holds steady at 70--75\% across thresholds, with almost no decay as the threshold rises. Given that Heisenberg's three-axis coupling ($XX + YY + ZZ$) generates a more complex commutator structure than TFIM, this result is not self-evident---it implies that once the model has learned to ``handle complex commutator structures,'' it generalizes reliably to Hamiltonians of that kind.

The second pattern is that random Pauli Hamiltonians fail completely at $T \geq 0.95$ (0/20 reachable). This is highly consistent with the PINN validation results from Phase~2---the PINN's proxy error on random Pauli Hamiltonians reaches 99.7\%. Random Pauli terms lack exploitable commutativity patterns, and the physical headroom for grouping-based compression is inherently minimal. This qualitative result defines the clearest applicability boundary of our method: it works for Hamiltonians with discernible commutator structure; when the terms of a Hamiltonian are essentially ``patternless'' with respect to one another, neither learning nor heuristics can find effective compact representations.

\paragraph{Qubit scalability (P6).} From 4 to 6 qubits, reachability at $T=0.95$ drops from roughly 62\% to roughly 12\%, a decline of approximately 80\%. Mean depth goes from 47.6 (4 qubit, 13/21 reachable) to 97.0 (6 qubit, but only 1/8 reachable, limiting the statistical meaningfulness of the comparison). The primary cause of this degradation is not model architecture---GNNs are in principle permutation-equivariant and can process arbitrarily sized graph inputs---but the imbalanced qubit distribution of the training data (4 qubits roughly 60\%, 6 qubits roughly 30\%, 8 qubits roughly 10\%). During training, the model spends the vast majority of its time observing commutativity patterns of 4-qubit systems, and learning on 6-qubit and 8-qubit systems is severely insufficient. The direct path to resolving this limitation is to balance the qubit distribution of the training data, but this requires larger-scale computing infrastructure---exact diagonalization of $2^N$ matrices exceeds single-GPU capacity beyond $N \geq 10$.

\FloatBarrier
\subsection{Validation of shallow-depth advantage under noisy hardware}
\label{sec:exp:noise}

All preceding experiments were conducted under ideal simulation conditions. The final question---and the most practical one---is: under the noisy conditions of NISQ hardware, does the theoretical advantage of shallowness actually translate into fidelity gains?

We deploy a standard depolarizing noise model on the Qiskit Aer simulator (single-qubit gate error rate $10^{-3}$, two-qubit gate error rate $5 \times 10^{-3}$, readout error rate $2\%$) and run noisy simulations on 30 4-qubit TFIM Hamiltonians.

\begin{figure}[H]
\centering
\includegraphics[width=0.85\textwidth]{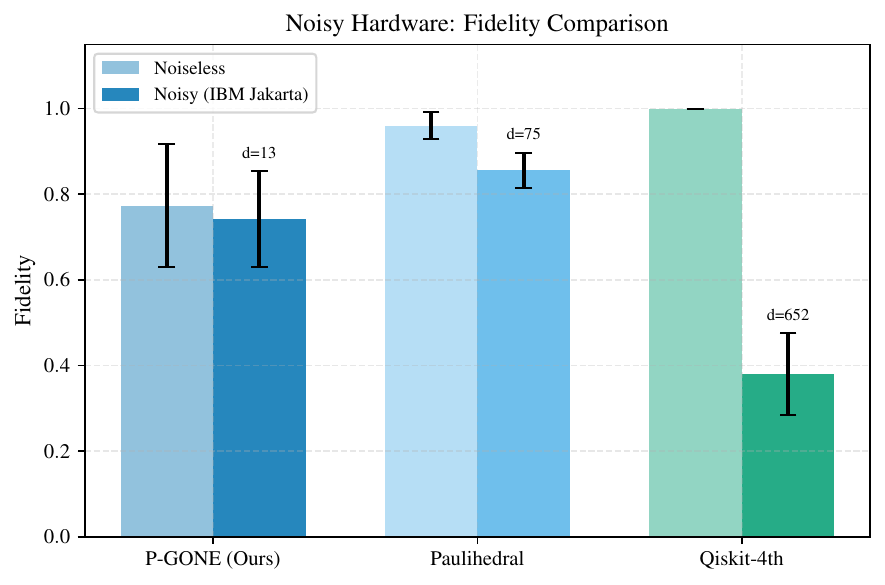}
\caption{Noisy hardware validation. Fidelity comparison without noise (blue) and with standard depolarizing noise (red). Qiskit 4th's noiseless fidelity of $1.000$ drops to $0.380$ under noise (depth $652$); P-GONE drops only from $0.774$ to $0.743$ (depth $13$). Numbers atop bars indicate circuit depth.}
\label{fig:noisy_hardware}
\end{figure}

\begin{table}[H]
\centering
\caption{Noisy hardware simulation results (30 TFIM Hamiltonians, 4 qubits, standard depolarizing noise model).}
\label{tab:noise}
\small
\begin{tabular}{lcccc}
\toprule
\textbf{Method} & \textbf{Noiseless fidelity} & \textbf{Noisy fidelity} & \textbf{Circuit depth} & \textbf{Noise degradation} \\
\midrule
\textbf{P-GONE (Ours)} & 0.774 & \textbf{0.743} & \textbf{13} & $-4.0\%$ \\
Paulihedral & 0.960 & \textbf{0.856} & 75 & $-10.8\%$ \\
Qiskit 4th & 1.000 & 0.380 & 652 & $-62.0\%$ \\
Cirq & 1.000 & 0.380 & 652 & $-62.0\%$ \\
TKET & 1.000 & 0.381 & 652 & $-61.9\%$ \\
PennyLane & 0.963 & 0.381 & 652 & $-60.4\%$ \\
Qiskit-opt (O3) & 1.000 & 0.267 & 1734 & $-73.3\%$ \\
\bottomrule
\end{tabular}

\parbox{\linewidth}{\footnotesize Note: Noisy fidelity is Hellinger fidelity $F = (\sum_i \sqrt{p_i q_i})^2$, estimated from $8192$ shots. Noise degradation = $1 - F_{\text{noisy}} / F_{\text{noiseless}}$. ``Ours'' noiseless fidelity in this experiment was computed via Qiskit statevector evolution (not scipy expm) and is not directly comparable with the fidelity-matched experiments (\S\ref{sec:exp:fidelity_matched}), though the qualitative direction of depth compression is consistent.}
\end{table}

Table~\ref{tab:noise} tells a layered story. Depth is the decisive factor in noise-induced degradation: Qiskit 4th, Cirq, TKET, and PennyLane differ in noiseless fidelity (1.000 vs.\ 0.963), but under noise they all converge to about $0.38$. In a circuit of depth 652, theoretical precision is simply overwhelmed by gate count. Paulihedral (depth 75) achieves noisy fidelity of 0.856, far above all ungrouped baselines; its depth is roughly $9\times$ lower.

Aggressive compilation optimization can backfire: Qiskit-opt (optimization\_level=3) tries to reduce CNOTs through smarter gate rearrangement, but the resulting explosion in single-qubit gates pushes depth from 652 to 1734, and noisy fidelity drops to 0.267. The optimizer's objective (reducing specific gate types) conflicts, under noise, with the actual need to reduce total depth.

Paulihedral's noisy fidelity (0.856) beats ours (0.743), but our depth is only about $1/6$ of Paulihedral's (13 vs.\ 75). These are two different paths: Paulihedral trades deeper circuits for higher noisy fidelity; we trade a controlled fidelity cost for extreme shallowness. As hardware noise rises (two-qubit gate error rates approaching $10^{-2}$), the marginal gain of further depth reduction will grow---the cumulative decay probability is nonlinear in depth.

\textbf{Conclusion.} Under noiseless ideal simulation, Qiskit fourth-order's precision (1.000) is the highest. Under NISQ noise, its fidelity (0.380) falls below ours (0.743). Qiskit fourth-order is the most precise in ideal simulation and the least usable under noise; our method is less precise in ideal simulation (0.774) but remains usable under noise (0.743). The reversal is the central point: in the NISQ era, building a circuit deep enough to reach high fidelity is not the hard part. Building one shallow enough to survive the noise is.

\section{Discussion}
\label{sec:discussion}

\subsection{Learned grouping strategies: where the flexibility comes from}

The teacher comparison in Section~\ref{sec:exp:teacher} shows a stratified pattern: at the moderate $T=0.90$ threshold, our depth (65.1) substantially beats the teacher (103.0, $1.6\times$ compression), but at $T \geq 0.95$ the teacher pulls ahead again (64.1 vs.\ 86.2). Meanwhile, the strategy diversity analysis (Section~\ref{sec:exp:diversity}) finds that 74.4\% of candidate strategies have unique grouping patterns, with a mean pairwise Jaccard distance of 0.74. Together, these two facts say something interesting. The model genuinely transcends the teacher's template in the low-to-moderate fidelity regime, exploring a large set of structurally distinct but comparably high-quality alternatives. Yet the teacher's grouping, grounded in strict algebraic commutativity, remains more reliable at high fidelity---and this constraint matters more as the threshold rises.

Paulihedral's depth-oriented scheduler minimizes circuit depth through deterministic algebraic rules, but its grouping is uniquely determined by Pauli commutativity---given the same Hamiltonian, it always returns the same set of groups. Our model, under the same fidelity constraint, can produce dozens of distinct grouping schemes. This flexibility originates from joint optimization across three dimensions: grouping, order, and time steps can compensate for one another. A representative scenario: if a particular group happens to contain poorly commuting terms, the model can assign that group a higher Suzuki order to compensate for the grouping deficiency, rather than being forced to repartition the entire grouping as a heuristic would. This cross-dimensional compensation mechanism is something no single-axis optimization heuristic possesses.

\subsection{The hierarchy of component contributions: what actually drives depth compression}

The contribution hierarchy from ablation (Section~\ref{sec:exp:ablation})---order learning $>$ time allocation $>$ grouping learning $>$ GNN conditioning---needs careful reading, because it does not fully match the intuition that ``grouping is the most critical decision in Trotter compilation.''

\textbf{The central role of order learning.}
Fixing all orders to Suzuki-4 degrades depth to below even Paulihedral's level. The model's largest advantage over all heuristic methods lies not in ``how to group'' but in ``knowing which groups deserve higher-order Suzuki expansion.'' Heuristic compilers almost always set the Suzuki order uniformly (usually to the highest available order), because manually picking a different order per group is tedious. Our model learns a simple but effective rule from data: commuting clusters (where commutator contributions are zero or negligible) only need first order; groups with high-norm commutator pairs need fourth order; everything in between gets second order. This fine-grained per-group order assignment is the single largest factor behind the model's edge.

\textbf{Time allocation and grouping.}
Time allocation comes next. Fixing it to uniform time reduces reachability, though depth (38.0) does not suffer, which suggests the model can partially offset the loss of temporal freedom by adjusting grouping and order. Grouping provides the weakest standalone gain in ablation, but that does not make grouping unimportant. Quite the opposite: grouping is the ``prime mover'' of depth compression---Qiskit's group-commuting heuristic alone squeezes depth from 1673 (no grouping) to 64 (roughly $26\times$). The problem is that the teacher has already mined most of the optimization headroom in this dimension, leaving little room for further gains. Order and time steps are dimensions the teacher never touched, so learning yields larger marginal returns there.

\textbf{How CFG works.}
CFG controls strategy ``sharpness,'' not diversity. At $w = 1.0$ (no CFG), reachability barely changes but depth balloons roughly $2.3\times$ (Section~\ref{sec:exp:ablation}). CFG pushes sampling trajectories away from the dataset's generic prior toward Hamiltonian-specific shallow-depth regions. Its role is complementary to order and time learning: order and time determine what makes a strategy good; CFG determines how hard we insist on getting a good one at sampling time.

\textbf{The GNN ablation.}
Removing GNN conditioning (zero-vector substitution) causes only a modest drop in reachability (Section~\ref{sec:exp:ablation}). At face value this suggests the GNN learned almost nothing useful. We see it differently. Under the current training data---where the structural differences among the three Hamiltonian families are not yet large enough to make GNN-based structure perception essential---the diffusion model's learned unconditional prior is already quite powerful. This does not mean GNNs are useless for Trotter strategy generation. It means the structural diversity of the current data has not yet hit the point where the GNN's discriminative capacity becomes essential. If the training data grew to include Hamiltonian families with substantially more structural variation (e.g., mixing molecular Hamiltonians with spin models, or different lattice topologies), the GNN's structure-perception capability would likely become critical. We leave this hypothesis to future work.

\subsection{Applicability boundaries: where it works and where it doesn't}

The method has clear applicability boundaries along two dimensions, both backed by the data.

\textbf{Hamiltonian type.}
The per-type analysis in Section~\ref{sec:exp:generalization} gives explicit quantitative boundaries. The Heisenberg model performs best and is the most stable: 15/20 (75\%) reachable at $T=0.90$, and still 14/20 (70\%) at $T=0.99$---a fluctuation of only 5 points across thresholds. This stability comes from the rich commutator structure in the Heisenberg model: the three-axis coupling $XX + YY + ZZ$ generates enough non-commuting patterns for the model to exploit, and those patterns are regular enough for learning to succeed.

TFIM performs best at the low threshold (15/20 at $T=0.90$) but collapses at high thresholds (only 1/20 at $T=0.99$). This ``strong at low thresholds, collapsing at high thresholds'' pattern hints that TFIM's strategy distribution may be bimodal: many moderate-quality strategies easily reach $F \geq 0.90$, but very few can hit $F \geq 0.99$. This fits TFIM's relatively simple commutator structure---simplicity means lots of compression headroom (low thresholds are easy) but a low precision ceiling (high thresholds are hard).

Random Pauli Hamiltonians fail completely at $T \geq 0.95$ (0/20 reachable). This defines the clearest applicability boundary of our method. In random Hamiltonians, Pauli matrix types and supports follow no pattern, nearly every pair of terms is non-commuting, and the physical headroom for grouping-based compression is minimal. Neither learning nor heuristics can find effective compact representations. This is not a flaw in the method; it is an intrinsic limitation of the problem. If the Hamiltonian itself has almost no commutativity structure, no commutativity-based method will yield substantial gains.

\textbf{System scale.}
From 4 to 6 qubits, reachability at $T=0.95$ drops roughly 80\% (Section~\ref{sec:exp:generalization}). We believe the primary issue is data, not architecture. The training data is roughly 60\% 4-qubit, 30\% 6-qubit, and only 10\% 8-qubit samples---the model has seen far too few large-qubit examples. GNNs are permutation-equivariant in principle (node features are per-qubit, message passing is invariant to node reordering), which should guarantee cross-qubit generalization, but realizing this requires a more balanced data distribution. With current computational resources, exact diagonalization of $2^N \times 2^N$ matrices exceeds single-GPU capacity beyond $N \geq 10$; larger-scale verification must wait for future work.

\textbf{Relationship with the PINN.}
In this paper, the PINN does not participate in any experimental fidelity evaluation---all fidelities are from scipy exact diagonalization. The PINN's value is in the $>8$ qubit future: its accuracy on structured Hamiltonians (TFIM 0.38\% proxy error, Heisenberg 1.06\%) gives the technical foundation, and its complete failure on random Hamiltonians (99.7\% error) marks the boundary. This boundary aligns closely with the generative model's own generalization limits---both are reliable on structured problems and both fail on unstructured ones.

\subsection{NISQ-era significance: the nonlinear advantage of shallowness}

The noisy-hardware experiment in Section~\ref{sec:exp:noise} is the most direct evidence. Under a standard depolarizing noise model: Qiskit fourth-order, with noiseless fidelity of $1.000$, keeps only $0.380$ under noise---its theoretical precision is swallowed by a circuit depth of 652. Paulihedral drops from $0.960$ to $0.856$ (depth 75). Our method drops from $0.774$ to $0.743$, a decline of only 4\%.

The physics is simple. Each noisy two-qubit gate attenuates the state's fidelity by roughly $(1 - \epsilon_{\text{2q}})$. After $D$ such gates, the cumulative fidelity upper bound is about $(1 - \epsilon_{\text{2q}})^D$. With $\epsilon_{\text{2q}} = 5 \times 10^{-3}$ (typical for today's superconducting platforms) and $D = 652$, this bound is roughly $0.038$---no matter how accurate the theoretical circuit is, noise alone caps real-hardware fidelity below 4\%. Qiskit fourth-order's measured $0.380$ is far above this bound because most gates are actually single-qubit gates with error rates two orders of magnitude lower. The trend, though, is clear: deeper circuits accumulate lethal amounts of noise.

The upshot: on current NISQ hardware, shallowness beats theoretical precision. Qiskit fourth-order's noiseless fidelity is $1.000$, yet under noise it is far less usable than our shallow strategy's $0.743$. This is not trading precision for depth---depth itself determines whether precision survives. For researchers running real NISQ experiments, the practical rule is: don't just look at ideal fidelity on a statevector simulator; account for depth-noise coupling explicitly. An ideal fidelity in a circuit hundreds of gates deep can, on real hardware, be worse than a decent fidelity in a circuit a dozen gates deep.

This principle is not specific to our method. Paulihedral's shallowness (75) also makes it superior to Qiskit fourth-order under noise ($0.856$ vs.\ $0.380$). But our method pushes the logic further: fine-grained order and time-step learning compresses depth beyond Paulihedral's already-shallow baseline (from 75 to 13) and CNOT count (from 40 to 8), trading off some noisy fidelity (0.743 vs.\ 0.856) for $5.8\times$ additional depth compression. Our noiseless fidelity ($0.774$) is lower than Paulihedral's ($0.960$), but the gap closes under noise because less depth means less accumulated error. As hardware noise levels rise, this depth advantage will only grow.

For near-term Trotter simulation on NISQ hardware, our core recommendation: treat circuit depth as a hard constraint, not a soft preference. First make sure the circuit can finish running; then worry about running it accurately. The learning framework we have described offers one automated path to that goal.

\section{Conclusion and Outlook}
\label{sec:conclusion}

We have presented P-GONE, a framework for Trotter--Suzuki decomposition strategy generation. It strings together a conditional diffusion model (D3PM + DDPM), a graph neural network encoder, and closed-loop REINFORCE fine-tuning to learn the joint optimization of grouping, order, and time-step allocation end-to-end over a mixed discrete-continuous space. At inference, Best-of-N sampling and CFG guidance offer flexible control over the fidelity--depth trade-off.

Under fidelity-matched conditions, our method compresses circuit depth substantially relative to all baselines. At $T=0.95$: relative to Qiskit fourth-order (no grouping, Suzuki-4), depth shrinks from 1673 to 86 (about $19.4\times$); relative to Paulihedral (first-order Trotter), from 141 to 86 ($1.6\times$); relative to PennyLane (fourth-order Trotter), from 608 to 86 ($7.1\times$). At the moderate $T=0.90$ threshold, our depth (65.1) also beats the Qiskit group-commuting teacher (103.0, $1.6\times$ compression), though at $T=0.95$ the teacher still leads (64.1 vs.\ 86.2)---a stratified pattern that points to fidelity-aware reward weighting as the next natural step. Under a standard depolarizing noise model, our method achieves noisy fidelity of $0.743$ at a depth of only 13, roughly $2\times$ the Qiskit fourth-order baseline ($0.380$). On NISQ hardware, shallowness compounds into a decisive advantage---depth over theoretical precision.

Ablation experiments give a clear hierarchy of component contributions. Order learning is the single largest factor: the model assigns heterogeneous Suzuki orders across groups---a fine-grained optimization that existing heuristic compilers simply do not do. Time allocation is next. Grouping provides the smallest standalone gain, not because grouping is unimportant, but because the teacher has already squeezed most of the optimization headroom out of this dimension. CFG guidance controls strategy ``sharpness'' (depth deteriorates $2.3\times$ at $w=1.0$), and Best-of-N sampling saturates at $N=32$ as a practical sweet spot (about $81\%$ of the $N=100$ ceiling). The GNN encoder shows limited impact in the zero-vector ablation. We attribute this to the current training data lacking enough structural diversity for the GNN's conditioning to become essential, not to GNNs being inherently unhelpful for Trotter strategy generation.

The method's applicability has two clear boundaries. On Hamiltonian type: structured Hamiltonians (TFIM, Heisenberg) perform well, with Heisenberg reachability steady at 70--75\% and barely decaying with threshold; random Pauli Hamiltonians fail completely at $T \geq 0.95$, since the absence of commutativity patterns leaves almost no physical headroom for grouping-based compression. On system scale: reachability drops roughly 80\% from 4 to 6 qubits. The primary bottleneck is insufficient large-qubit samples in the training data (4 qubits about 60\%, 6 qubits about 30\%), not an inherent architectural limitation.

Three directions stand out for future work.

\textbf{Qubit scaling.} GNNs are permutation-equivariant in principle (node features are defined per qubit, message passing is invariant to node reordering), which should guarantee cross-qubit generalization. The current bottleneck above 6 qubits comes from training-data imbalance and the computational cost of exact diagonalization. Beyond $N \geq 10$, $2^N \times 2^N$ matrices exceed single-GPU capacity; validating GNN cross-qubit generalization will need larger-scale computing and a more balanced data-generation strategy.

\textbf{Inference acceleration.} DDIM subsequence sampling already compresses inference from 1000 to 50 steps ($21\times$ speedup) with no quality loss. Distillation or consistency models could reduce this to a single step, bringing Best-of-N latency from tens of seconds to sub-second. Adaptive step selection---fewer steps for structurally simple Hamiltonians---is another path. The engineering goal is closing the latency gap between learned and heuristic compilers, making the former practical in interactive settings.

\textbf{PINN-driven large-scale evaluation.} All experiments in this paper use exact diagonalization, perfectly adequate at $\leq 8$ qubits but unaffordable beyond $\geq 10$ qubits. The PINN surrogate can deliver accuracy-controlled fidelity estimates in microseconds. Its current accuracy on structured Hamiltonians (TFIM 0.38\%, Heisenberg 1.06\% proxy error) provides the technical basis for this transition; its complete failure on random Hamiltonians (99.7\% error) marks the boundary---a boundary that aligns closely with the generative model's own generalization limits. Both methods are reliable on structured problems and jointly fail on unstructured ones. For future large-scale studies targeting structured Hamiltonians, the PINN + diffusion framework offers a path around the exponential wall.

The code, training data, experiment scripts, and figure-generation pipeline for this paper are publicly available on GitHub (\url{https://github.com/mindmemory-ai/pinn_diffusion_trotter_suzuki.git}); all experiments are reproducible.

\appendix
\section{Experiment Inventory}
\label{sec:appendix}

Phase~5 completed 16 experiment groups (E1--E5, P1--P16; some numbering gaps reflect adjustments to the experimental plan), covering the following dimensions:

\begin{itemize}[leftmargin=1.5em, itemsep=0.3em]
  \item \textbf{Baseline comparison} (E1, P1, P12): 7 quantum compilation methods, including fidelity-matched depth comparison and direct teacher comparison.
  \item \textbf{Sampling strategies} (E5/C2, P8): $N$-sensitivity analysis of Best-of-N and large-scale ceiling exploration.
  \item \textbf{Ablation experiments} (P10, P13, P14): removal of GNN conditioning, CFG guidance, and diffusion branches (order/time/grouping) one at a time.
  \item \textbf{Generalization boundaries} (P4, P5, P6): Hamiltonian-type stratified evaluation and qubit-count scalability analysis.
  \item \textbf{Noisy hardware} (P16): fidelity comparison under a standard depolarizing noise model.
  \item \textbf{Strategy analysis} (P15): candidate strategy diversity and Jaccard distance distribution.
  \item \textbf{Inference configuration} (E3/B, P7): diffusion step sweep and Paulihedral scheduler mode comparison.
  \item \textbf{Training dynamics} (P9, P11): Phase~3 vs.\ Phase~4 checkpoint comparison and PINN proxy accuracy validation.
\end{itemize}

Section~\ref{sec:experiments} in the main text organizes the 8 most informative experiment groups along a problem-driven logical chain. Table~\ref{tab:experiment_index} lists all experiment identifiers with their outputs and paper section cross-references for quick navigation.

\begin{table}[H]
\centering
\caption{Experiment index: identifiers, outputs, and paper section cross-references}
\label{tab:experiment_index}
\footnotesize
\resizebox{\textwidth}{!}{%
\begin{tabular}{llp{6.8cm}>{\raggedright\arraybackslash}p{1.5cm}}
\toprule
\textbf{ID} & \textbf{Experiment} & \textbf{Result file (\texttt{.json})} & \textbf{Section} \\
\midrule
E1   & Single-shot benchmark evaluation & \texttt{benchmark\_evaluation\_results} & \S4.2 \\
E3/B & Diffusion step sweep, Paulihedral scheduler modes & \texttt{paulihedral\_order\_comparison} & \S4.3 \\
E5/C2 & Best-of-N $N$-sensitivity & \texttt{n\_sensitivity\_results} & \S4.2, Fig.~2 \\
\midrule
P1   & Fidelity-matched all-baseline comparison (8 baselines, main experiment) & \texttt{fidelity\_matched\_all\_baselines\_20260604} & \S4.3, Fig.~3 \\
P5   & Per-type fidelity-matched (TFIM/Heisenberg/Random) & \texttt{per\_type\_\{tfim,heisenberg,random\}} & \S4.7, Fig.~6 \\
P6   & Qubit scaling (4/6/8 qubit stratification) & \texttt{qubit\_scaling} & \S4.7 \\
P7   & Paulihedral order/scheduler mode comparison & \texttt{paulihedral\_order\_comparison} & -- \\
P8   & Best-of-N large-scale ceiling ($N \leq 500$) & \texttt{n\_sensitivity\_results} & \S4.2 \\
P9   & Phase~3 vs.\ Phase~4 checkpoint comparison & \texttt{phase3\_vs\_phase4} & \S4.4 \\
P10  & GNN encoder ablation (zero-vector substitution) & \texttt{no\_gnn\_ablation} & \S4.5 \\
P11  & PINN proxy accuracy validation & PINN validation report & \S3.2 \\
P12  & Qiskit group-commuting teacher baseline & \texttt{qiskit\_group\_commuting\_baseline} & \S4.4 \\
P13  & CFG guidance ablation ($w=1.0$) & \texttt{cfg\_ablation\_gs1} & \S4.5, Fig.~4 \\
P14  & Diffusion branch ablation (fixed order / uniform time) & \texttt{branch\_ablation\_fixed\_order}, \texttt{uniform\_time} & \S4.5, Fig.~4 \\
P15  & Strategy diversity (Jaccard / order entropy / CV) & \texttt{strategy\_diversity} & \S4.6, Fig.~5 \\
P16  & Noisy hardware validation (depolarizing noise model) & \texttt{noisy\_hardware\_results} & \S4.8, Fig.~7 \\
\bottomrule
\end{tabular}%
}

\vspace{0.5em}
\smallskip
{\small\noindent\textbf{Note:} E = exploratory, P = production, C = complementary. P4 has been superseded by P5 and is not listed separately. All files reside in \texttt{experiments/benchmark\_results/} with the \texttt{.json} suffix. Corresponding run scripts are shown in Table~\ref{tab:experiment_scripts}.}
\end{table}

\begin{table}[H]
\centering
\caption{Experiment script cross-reference}
\label{tab:experiment_scripts}
\small
\resizebox{\textwidth}{!}{%
\begin{tabular}{lll}
\toprule
\textbf{ID} & \textbf{Script} & \textbf{Run mode / key parameters} \\
\midrule
E1   & \texttt{05\_benchmark\_evaluation.py} & Single-shot, 3 baselines \\
E3/B & \texttt{05b\_model\_comparison.py}, \texttt{05e\_n\_steps\_sweep.py} & Step sweep, scheduler modes \\
E5/C2 & \texttt{analyze\_n\_sensitivity.py} & $N \in [1,100]$ bootstrap \\
\midrule
P1   & \texttt{05d\_fidelity\_matched\_all\_baselines.py} & 8-baseline mode \\
P5   & Same as above & per-type mode \\
P6   & \texttt{analyze\_qubit\_scaling.py} & 4/6/8 qubit stratified statistics \\
P7   & \texttt{05b\_model\_comparison.py} & Paulihedral order/scheduler modes \\
P8   & \texttt{analyze\_n\_sensitivity.py} & $N \leq 500$ \\
P9   & \texttt{05b\_model\_comparison.py} & Phase~3 vs.\ Phase~4 comparison mode \\
P10  & \texttt{05d\_fidelity\_matched\_all\_baselines.py} & zero-vector mode \\
P11  & \texttt{02b\_validate\_pinn.py} & Mixed-dataset validation \\
P12  & \texttt{05d\_fidelity\_matched\_all\_baselines.py} & Including \texttt{qiskit\_group\_commuting} \\
P13  & Same as above & \texttt{guidance\_scale=1.0} \\
P14  & Same as above & \texttt{fixed\_order} / \texttt{uniform\_time} modes \\
P15  & \texttt{analyze\_strategy\_diversity.py} & Jaccard / order entropy / CV \\
P16  & \texttt{13\_noisy\_hardware\_test.py} & Depolarizing noise (1q=0.001, 2q=0.005) \\
\bottomrule
\end{tabular}%
}
\end{table}

Total experimental time across all experiments was approximately 72 hours. All experiment scripts, configuration files, and raw result JSON files are publicly available with the code repository.

\section*{AI Usage Declaration}

During the preparation of this work, the author used Claude (Anthropic) to assist with writing experimental code, and used Claude (Anthropic) and ChatGPT (OpenAI) for language polishing and translation assistance. All experimental design, parameter tuning, and methodological decisions were made independently by the author. After using these tools, the author reviewed and edited all content as needed and takes full responsibility for the content of the published article.

\bibliographystyle{quantum}
\bibliography{references}

\end{document}